\documentclass[aps,prb,superscriptaddress,twocolumn,showpacs,floatfix]{revtex4}

\usepackage{amsmath,amsfonts,bm,graphicx,hyperref,hypernat}

\newcommand{\eps}{\varepsilon}
\newcommand{\om}{\omega}
\newcommand{\proj}[2]{|#1\rangle\!\langle #2|}
\newcommand{\norm}[1]{|\! |#1|\! |_2}

\begin{document}

\title{Current noise in a vibrating quantum dot array}

\author{Christian Flindt}
\email{cf@mic.dtu.dk}

\affiliation{MIC - Department of Micro and Nanotechnology,
             Technical University of Denmark,
             DTU - Building 345east,
             DK-2800 Kongens Lyngby, Denmark}

\author{Tom\'a\v s Novotn\'y}
\email{tno@mic.dtu.dk}

\affiliation{MIC - Department of Micro and Nanotechnology,
             Technical University of Denmark,
             DTU - Building 345east,
             DK-2800 Kongens Lyngby, Denmark}

\affiliation{Department of Electronic Structures,
     Faculty of Mathematics and Physics, Charles University,
     Ke Karlovu 5, 121 16 Prague, Czech Republic}

\author{Antti--Pekka Jauho}
\email{antti@mic.dtu.dk}

\affiliation{MIC - Department of Micro and Nanotechnology,
             Technical University of Denmark,
             DTU - Building 345east,
             DK-2800 Kongens Lyngby, Denmark}

\date{\today}

\begin{abstract}
We develop methods for calculating the zero-frequency noise for
quantum shuttles, i.e. nanoelectromechanical devices where the
mechanical motion is quantized.  As a model system we consider a
three-dot array, where the internal electronic coherence both
complicates and enriches the physics. Two different formulations
are presented: (i) quantum regression theorem, and (ii) the
counting variable approach. It is demonstrated, both analytically
and numerically, that the two formulations yield identical
results, when the conditions of their respective applicability are
fulfilled. We describe the results of extensive numerical
calculations for current and current noise (Fano factor), based on
a solution of a Markovian generalized master equation. The results
for the current and noise are further analyzed in terms of Wigner
functions, which help to distinguish different transport regimes
(in particular, shuttling vs.\ cotunneling). In the case of weak
inter-dot coupling, the electron transport proceeds via sequential
tunneling between neighboring dots. A simple rate equation with
the rates calculated analytically from the $P(E)$-theory is
developed and shown to agree with the full numerics.
\end{abstract}


\maketitle

\section{Introduction}

As the advances of the technology push the size of the electronic
components towards the atomic scale new interesting phenomena
influencing the electronic transport emerge. New research fields,
e.g.\ molecular electronics, spintronics, or nanoelectromechanical
systems (NEMS) have appeared.  A common theme is the combination
of quantum transport and a subtle interplay between various
degrees of freedom which plays an essential role for the
functionality of the device. This paper focuses on the
NEMS,\cite{cra-sci-00,cleland,ble-phr-04} a logical extension of
the now established technology of MEMS, where the electronic (or
magnetic) degrees of freedom are coupled to a mechanical degree of
freedom. While still in its infancy, NEMS have already attracted
much attention both experimentally \cite{cle-nat-98,
par-nat-00,erb-prl-01,bra-preprint-03,kno-nat-03,lah-sci-04} and
theoretically.\cite{gor-prl-98, wei-epl-99, boe-epl-01,
pol-pra-01, nis-prb-01, fed-epl-02, arm-prb-02, moz-prl-02,
nis-prl-02, smi-prb-03, bra-prb-03, mcc-prb-03, nov-prl-03,
fle-prb-03,braig-prb-03,mit-preprint-03,wer-epl-04,bra-preprint-04,
pis-preprint-04,arm-preprint-04,cht-preprint-04,isa-preprint-04,
smi-prb-04,fed-prl-04,nov-preprint-04,bla-preprint-04}

A measurement of the stationary IV-characteristic of a NEMS device
does not always yield enough information to uniquely identify the
underlying microscopic charge transport mechanism. A point in case
is  the C$_{60}$ SET experiment by Park et al.\cite{par-nat-00}
where two alternative interpretations, namely incoherent phonon
assisted tunneling\cite{boe-epl-01, mcc-prb-03, fle-prb-03,
braig-prb-03} or shuttling,\cite{gor-prl-98, fed-epl-02} are
plausible. The current noise provides another important
characteristics, supplementary to the mean
current.\cite{kogan,bla-phr-00,bee-pht-03} The Fano factor, being
the ratio between the zero-frequency component of the noise
spectrum and the mean current, characterizes the degree of
correlation between charge transport events and is a powerful
diagnostic tool which helps to distinguish various transport
mechanisms possibly resulting in the same mean current. Therefore,
studies of the current noise in NEMS have formed an active field
of research.\cite{mit-preprint-03,pis-preprint-04,arm-preprint-04,
cht-preprint-04,isa-preprint-04,nov-preprint-04,bla-preprint-04}
These studies considered noise in movable singe-electron
transistors in a number of different configurations.

To the best of our knowledge, the effects of internal coherence of
the electronic subsystem on the noise in NEMS have not been
considered so far. The coherence is not a dominating feature in a
system consisting of a single-level molecule or quantum dot.
However, in a setup consisting of an array of dots the role of the
electronic coherence within the array is of central importance.
Its influence on the current in {\em static} quantum dot arrays
has been studied intensively
\cite{mid-prl-93,sta-prl-94,sto-prb-96,gur-prb-96} and, more
recently, also on the noise.\cite{ela-pla-02} Also, the mean
current dependence on various system parameters in movable quantum
dot arrays has already been studied.\cite{arm-prb-02,bra-prb-03}
Thus, the study of noise in a movable quantum dot array is the
central theme in this work.

Specifically, we study an array of three quantum dots in the
strong Coulomb blockade regime with a movable central dot. This
model was proposed as a quantum shuttle by Armour and
MacKinnon\cite{arm-prb-02} extending the original one-dot shuttle
proposal by Gorelik et al.\cite{gor-prl-98} The electronic
coherence within the array combined with the mechanical degree of
freedom changes qualitatively the transport through the array as
compared to both a static array or a one-dot SET-NEMS. In
particular, there are two competing electron transfer mechanisms
through the array: either sequential tunneling or cotunneling
(virtual transition) via the central dot. The state of the
oscillator further influences these two basic mechanisms which
leads to a possibility of many different transport regimes
depending sensitively on the interplay of the parameters of the
model. Roughly speaking, as we shall see cotunneling is associated
with super-Poissonian values of the Fano factor (sometimes as high
as $\approx 50$) while the sequential tunneling is accompanied by
sub-Poissonian Fano factors.\cite{suk-prb-01} Similar conclusions
have been reported in recent literature for different but related
systems, and a detailed discussion is given in sections to follow.

We have recently published two Letters on quantum shuttles,
\cite{nov-prl-03,nov-preprint-04} and while the present paper
addresses a somewhat different physical system, it makes heavy use
of the techniques developed in the two Letters. Since we believe
that the techniques may have a wide range of applications, we use
this opportunity to describe our general approach to quantum
shuttles and expose the theoretical machinery in more detail. The
paper is organized as follows. In Section 2 we introduce our model
of the three-dot quantum shuttle which is quite similar to the one
considered in Ref.~\onlinecite{arm-prb-02}. The total Hamiltonian
consisting of the ``system" (both mechanical and electronic
degrees of freedom of the quantum dot array), the leads, and a
generic heat bath is used to illustrate the derivation of a
description based on Markovian generalized master equation which
was the starting point of Ref.~\onlinecite{arm-prb-02}. Along the
way from the Hamiltonian to the generalized master equation we
identify several tacit assumptions used in previous studies
(including ours) and point out several issues of potential
importance not addressed so far within the field of NEMS. While we
are not able to resolve all of these issues we believe that
spelling them out is an important first step towards their
solution. In particular, we address the problem of the assumed
additivity of two kinds of baths acting on the system (the Fermi
seas of the leads  and the heat bath weakly coupled to the
system). Another point of concern is the possible spurious
breaking of the charge conservation by the weak-coupling
prescription between the heat bath and the system with internal
coherence. We close Sec.~2 with a short introduction to the
superoperator formalism.

In Section 3 we develop the theory of the zero-frequency component
of the current noise spectrum for a  NEMS  device where the
electron transfer between the system and the leads is described by
a classical Markov process, i.e.\ in the wide band approximation
and high bias limit. We present two methods of the evaluation of
the noise spectra. If the whole system dynamics can be described
by a Markovian generalized master equation we can use the quantum
regression theorem. The other method relies on the counting
variable approach and calculates the zero-frequency current noise
as the charge diffusion coefficient across a given junction
between the system and a lead. As we show further in Sec.~3 the
two approaches yield equivalent results provided that the dynamics
of the system is (quantum) Markovian and that charge conserving
approximations are used. We finish Sec.~3 by a qualitative
discussion of the numerical evaluation of the noise spectra. This
is a non-trivial task due to large dimensions of the involved
matrices. Further details of the numerical algorithm (Arnoldi
iteration and generalized minimum residual method) are given in
Appendix A.

We present the results of our numerical and analytical
calculations in Section 4. Generic features observed in the
numerical curves are interpreted phenomenologically. Next, we
study different limiting cases. The first limit is that of small
damping which is relevant for shuttling accompanied by relatively
small Fano factors (down to $\approx 0.25$) and strong inelastic
cotunneling accompanied by huge Fano factors. These two mechanisms
may coexist leading to a dramatic dependence of the Fano factor on
parameters as the relative weight of the two mechanisms is
changed. The second limit considered is the limit of weak coupling
between adjacent dots which leads to sequential tunneling assisted
by an equilibrated oscillator, at least in a certain range of
other parameters. In the sequential tunneling limit we fully
reproduce the numerical results with (semi-)analytic
rate-equation-based theory with the rates determined by the
standard $P(E)$-theory as functions of the model parameters. The
technical details of the analytic calculations are sketched in
Appendix B. We state our conclusions in Sec.~5.

\section{Three-dot quantum dot array}
\subsection{The Model}

\begin{figure}
  \centering
  \includegraphics[width=85mm]{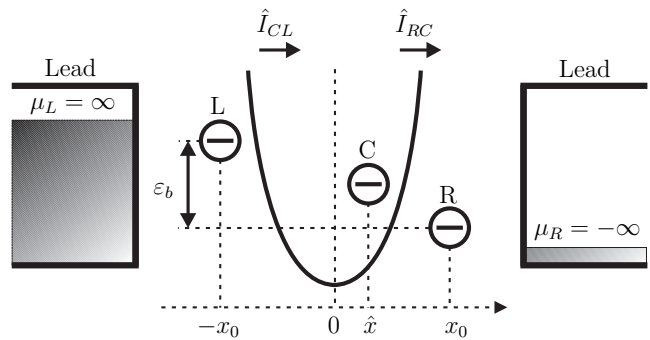}
  \caption{Schematic picture of the three dot system. The outer
  dots are fixed --- the left one (L) at the position $-x_0$ and the right one
  (R) at $x_0$, while the central one (C) can move (position $\hat{x}$) in
  a harmonic confining potential. It also interacts with a heat bath causing damping
  and thermal noise. The outer dots whose respective energy levels are
  de-aligned by the device bias ($\eps_b$) are coupled to the full or empty electronic
  reservoirs (leads), respectively. The current flows within the system due to tunneling
  between the left and central dot and the central and right dot and is described by the
  corresponding current operators $\hat{I}_{CL},\, \hat{I}_{RC}$.}\label{setup}
\end{figure}

Armour and MacKinnon \cite{arm-prb-02} introduced a model of a
three-dot array whose central dot is movable. The array is assumed
to be in the strong Coulomb blockade regime in which only two
charge states (none or one extra electron which we refer to as
unoccupied or singly occupied) of the whole array, separated by an
energy difference $\eps_0$, are allowed in the considered bias
range. This can be achieved by a suitable gating of the array
which makes the two charge states energetically close while a very
high charging energy prohibits addition or removal of other
electrons to/from the array. The array is coupled to two leads
with a high bias applied between them. The bias is smaller than
the charging energy for addition or removal of other electrons but
otherwise it is the largest energy scale in the model.

The moving central dot interacts with its surroundings and the
dissipative dynamics is described by the interaction with a
generic heat bath. We modify the original model slightly in that
we do not consider the additional hard wall potential at the
position of the outer dots $\pm x_0$ employed by Armour and
MacKinnon \cite{arm-prb-02} so that the central dot moves in a
strictly harmonic potential in our case (see Fig.~\ref{setup}).
While the hard wall potential is physically well motivated it
complicates the numerical treatment and we believe that it does
not have any significant impact on the nature of our results.
Therefore, in our model the amplitude of oscillations in some
regimes can exceed $x_0$. The hard wall potential can be
straightforwardly incorporated in our formalism.  It should be
noted, however, that the various models for dissipation used in
the literature, and also adopted in our work, are best justified
for the pure harmonic potential. Also, as in
Ref.~\onlinecite{arm-prb-02}, we consider spinless electrons.

The Hamiltonian reads
\begin{equation}\label{Hamiltonian}
\begin{split}
    \hat{H} &= \hat{H}_{\rm osc} + \hat{H}_{\rm el} + \hat{H}_{\rm el-osc} +
               \hat{H}_{\rm leads} + \hat{H}_{\rm el-leads}\\
            &+ \hat{H}_{\rm bath}+ \hat{H}_{\rm osc-bath} + \hat{H}_{\rm CT}
\end{split}
\end{equation}
where
\begin{equation}
       \hat{H}_{\rm osc}=\frac{\hat{p}^2}{2m}+\frac{m\om_0^2\hat{x}^2}{2}
       \tag{1a}
\end{equation}
describes the mechanical center-of-mass motion of the central dot
as a one-dimensional harmonic oscillator with mass $m$ and
frequency $\om_0$. The next two terms specify the electronic
structure of the array in the strong Coulomb blockade regime
(i.e.\ no double occupancy in the whole array  --- the vectors
$|I\rangle$ with $I=0,L,C,R$ span its entire electronic Hilbert
space) and the electromechanical coupling within the array
\begin{equation}
\begin{split}
     & \hat{H}_{\rm el} + \hat{H}_{\rm el-osc} = \frac{\eps_b}{2}\proj{L}{L}
      - \frac{\eps_b}{2}\proj{R}{R} +\eps_0\proj{0}{0}\\
     & + t_L(\hat{x})\Big(\proj{L}{C}+\proj{C}{L}\Big)
      + t_R(\hat{x})\Big(\proj{C}{R}+\proj{R}{C}\Big)\\
      &- \frac{\eps_b}{2x_0}\,\hat{x} \proj{C}{C}
\end{split}\tag{1b}
\end{equation}
with $t_L(\hat{x})=-V_0 e^{-\alpha (x_0+\hat{x})},\,
t_R(\hat{x})=-V_0 e^{\alpha (\hat{x}-x_0)}$. We associate the
energies $\tfrac{\eps_b}{2},-\tfrac{\eps_b}{2}$, and $\eps_0$ with
the left and right dot and the empty array, respectively, while
the energy level of the central dot is chosen as the reference
energy, and hence put to zero. The device bias $\eps_b$ is the
difference between the energy of the left and the right dot (which
can be induced by suitable gating of the different dots) and
$2x_0$ is the distance between the two outer dots. The terms
proportional to $t_{L,R}(\hat{x})$ describe a position-dependent
hopping between the left and central or central and right dots
enabling the tunneling current to flow through the array. These
terms contribute both to the static part of the Hamiltonian
(zeroth order in $\hat{x}$) as well as to the electromechanical
coupling. The parameter $\alpha$ equals the inverse tunneling
length and determines the strength of the exponential
$\hat{x}$-dependence of the hopping elements which may lead to the
shuttling instability.\cite{gor-prl-98,arm-prb-02,nov-prl-03} The
last term gives the electromechanical coupling due to the
electrostatic force acting on the oscillator when the central dot
is charged.

The outer dots of the array are assumed to couple via standard
tunneling terms to two non-interacting leads:
\begin{equation}
\begin{split}
    \hat{H}_{\rm leads} &+ \hat{H}_{\rm el-leads} = \sum_{k;\beta=L,R}
    \eps_{k\beta}\hat{c}_{k\beta}^{\dag}\hat{c}_{k\beta}\\
    & + \sum_{k;\beta=L,R} V_{k\beta}\Bigl(\hat{c}_{k\beta}^{\dag}\proj{0}{\beta}
       + \proj{\beta}{0}\hat{c}_{k\beta}\Bigr)\ .
\end{split}\tag{1c}
\end{equation}
The leads are held at different electrochemical potentials
$\mu_{L,R}$ whose difference gives the bias across the array. We
assume that the tunneling densities of states
$\Gamma_{\beta}(\eps)=\tfrac{2\pi}{\hbar}\sum_k
|V_{k\beta}|^2\delta(\eps-\eps_{k\beta})$ are energy-independent
(and equal, just for convenience), i.e.\
$\Gamma_{\beta}(\eps)=\Gamma$, known as the wide-band limit. It is
necessary for the so-called {\em first Markov
approximation},\cite{gardiner,gar-preprint-03} used later on, to
hold. Further, we assume $\mu_L\to\infty,\,\mu_R\to -\infty$.
These assumptions are necessary for the derivation of the
Markovian dynamics of the array.

Finally, we introduce a generic heat bath consisting of an
infinite set of harmonic oscillators linearly coupled to the
position of the central dot (Caldeira-Leggett model\cite{weiss})
which simulates the dissipative interaction of the center-of-mass
motion of the central dot with its environment
\begin{equation}
\begin{split}
\hat{H}_{\rm bath}+ \hat{H}_{\rm osc-bath} + \hat{H}_{\rm CT}&=
     \sum_j \Bigl(\frac{\hat{p}_j^2}{2m_j}+\frac{m_j \om_j^2 \hat{x}_j^2}{2}\Bigr) \\
     &- \sum_j c_j \hat{x}_j
     \hat{x}-\frac{m}{2}\Delta\om^2\hat{x}^2\ .
\end{split}\tag{1d}
\end{equation}
The bath is characterized by its spectral density
$J(\om)=\tfrac{\pi}{2}\sum_j\tfrac{c_j^2}{m_j\om_j}\,\delta(\om-\om_j)$.
We take it in the Ohmic form\cite{weiss} $J(\om)= m\gamma\om
f(\tfrac{\om}{\om_c})$ where we have introduced the damping
coefficient $\gamma$ and $f(\tfrac{\om}{\om_c})$ is a model
specific cut-off function $f(x\to 0)\to 1$. As long as the cut-off
frequency is much bigger than the frequency of the oscillator
($\om_c\gg\om_0$) $f$ would only contribute to the renormalization
of $\om_0^2\to\om_0^2+\Delta\om^2$ with
$\Delta\om^2=-\tfrac{1}{m}\sum_j\tfrac{c_j^2}{m_j\om_j^2}=
-\tfrac{2}{\pi}\int_0^{\infty}d\om\tfrac{J(\om)}{m\om}
=-\tfrac{2\gamma}{\pi}\int_0^{\infty}d\om f(\tfrac{\om}{\om_c})$.
Here, we have explicitly included the standard counter-term
$\hat{H}_{\rm CT}$ cancelling this renormalization so that the
bath solely induces dissipation and the cut-off function can be
replaced by unity.

\subsection{Generalized Master Equation}
\label{GMEsection}

For the description of the model we use the language of quantum
dissipative systems.\cite{weiss} As the ``system" (or ``device")
we take the electronic states of the dots in the array (including
the unoccupied state) plus the one-dimensional oscillator
describing the center-of-mass motion of the central dot. The
electronic leads coupled to the outer dots and the heat bath
interacting with the center-of-mass degree of freedom of the
central dot constitute the reservoirs. The Hamiltonian of the
system is then $\hat{H}_0=\hat{H}_{\rm osc} + \hat{H}_{\rm el} +
\hat{H}_{\rm el-osc}$. For further reference we also introduce the
Hamiltonian of all mechanical degrees of freedom, i.e.\ of the
oscillator and the bath, reading
$\hat{H}_{\mathrm{osc}}'=\hat{H}_{\mathrm{osc}}+\hat{H}_\mathrm{osc-bath}
+\hat{H}_{\mathrm{bath}}+\hat{H}_{\rm CT}$. The task is now to
integrate out the degrees of freedom of the reservoirs to end up
with an equation of motion for the system density operator. We
outline how the derivation proceeds in two steps first integrating
out the leads in the high bias limit and then the heat bath in the
weak coupling limit to get a generalized master equation (GME) for
the system density operator.

As in previous papers,\cite{arm-prb-02, nov-prl-03, modena} we
work in the high bias limit in which the bias between the leads is
much higher than any other involved energy scale but the charging
energy (cf.\ Ref.~\onlinecite{gur-prb-96} and Fig.~\ref{setup}).
The high bias assumption together with the wide-band limit means
that after integrating out the leads the resulting dynamics of the
system and heat bath is still Markovian. Following the derivation
by Gurvitz and Prager\cite{gur-prb-96} one can obtain the
equations of motion for the density matrices
$\hat{\sigma}^{(n)}(t)$ of the system plus heat bath resolved with
respect to the number of electrons $n$ which have tunneled to the
right lead by time $t$. We use the block notation analogous to the
one used in Ref.~\onlinecite{arm-prb-02} ($\hbar=1$ throughout the
paper except for figures):
\begin{equation}\label{GME1}
\begin{split}
\dot{\hat{\sigma}}^{(n)}_{00}&=-i[\hat{H}'_{\rm
osc},\hat{\sigma}_{00}^{(n)}] - \Gamma\hat{\sigma}_{00}^{(n)}
+\Gamma\hat{\sigma}_{RR}^{(n-1)},\ n=0,1,\dots\\
\dot{\hat{\sigma}}^{(n)}_{IJ}&=-i\langle I|[\hat{H}_{\rm el} +
\hat{H}'_{\rm osc}+\hat{H}_{\rm
el-osc},\hat{\sigma}^{(n)}]|J\rangle \\
&\quad + \langle I|\mathcal{K}_{\rm driv}
\hat{\sigma}^{(n)}|J\rangle\ \text{ for } I,J=L,C,R\ .
\end{split}
\end{equation}
Here $\hat{\sigma}_{IJ}=\langle I|\hat{\sigma}|J\rangle$ are still
operators in the oscillator and bath space. The ``driving" kernel
$\mathcal{K}_{\rm driv}$ due to the coupling to the leads acts
non-trivially only on the electronic degrees of freedom and as
unity on all the others. Hence also it can be written in the block
notation
\begin{equation}\label{driving}
 \mathcal{K}_{\rm driv}\hat{\sigma}=\Gamma\begin{pmatrix}
   \hat{\sigma}_{00} & 0     & -\hat{\sigma}_{LR}/2 \\
   0    & 0     & -\hat{\sigma}_{CR}/2 \\
   -\hat{\sigma}_{RL}/2 & -\hat{\sigma}_{RC}/2  & -\hat{\sigma}_{RR}   \\
\end{pmatrix}
\end{equation}
where the tunneling density of states $\Gamma$ describes the
injection rate from/to the leads. We still have to consider the
off-diagonal block elements of the density matrix
$\hat{\sigma}_{0I},\,\hat{\sigma}_{I0}$ with $I=L,C,R$. They
describe coherences between system states containing different
number of electrons. In the formalism by Gurvitz and Prager
\cite{gur-prb-96} these off-diagonal elements are identically zero
by the construction of the theory (see also
Ref.~\onlinecite{arm-prb-02}). In other works, e.g.\ in
Ref.~\onlinecite{gar-preprint-03}, they can in principle appear,
at least indirectly. In any case, whatever method is applied to
our system, they are always decoupled from the rest of the
elements. Moreover, they do not enter any expressions for
quantities of physical interest that we consider,  and can
therefore be discarded.

The GME for $\hat{\sigma}(t)=\sum_n\hat{\sigma}^{(n)}$ is found by
summing \eqref{GME1} over $n$ with the boundary
condition\cite{gur-prb-96} $\hat{\sigma}^{(-1)}\equiv 0$. Due to
this boundary condition the GME for $\hat{\sigma}(t)$ is formally
the same as \eqref{GME1} just with the superscript index $(n)$
omitted. This GME is used in subsection \ref{seq_tun} and Appendix
\ref{P(E)-theory} in the sequential tunneling limit  to derive a
rate equation, from which both current and noise can be
calculated, and compared to the full numerical evaluation.

In general, there is no simple approximative analytic treatment of
the problem nor is a direct numerical solution possible due to the
presence of the infinite number of bath degrees of freedom as a
part of the system. To proceed we have to integrate out the bath
degrees of freedom to be left with the electronic and oscillator
degrees of freedom only which can be handled numerically. This
could in principle be done in the weak coupling limit between the
device and the heat bath by a perturbation expansion in the
$c_j$'s. This would amount to finding the ``free" evolution of the
device first, i.e.\ the evolution without the coupling to the heat
bath but with coupling to the leads included. However, this free
evolution is not unitary which significantly hinders any attempt
to proceed. Even in the case of small coupling $\Gamma$ to the
leads, when the driving Liouvillean is neglected\footnote{The
influence of two ``baths" on a single system may lead to
interesting physical phenomena --- see, e.g.,
Ref.~\onlinecite{gur-prl-03} where the subtle interplay between a
detector and bath is studied. Similar study in the context of NEMS
has not been carried out yet.}, one should diagonalize the device
Hamiltonian (including the electromechanical coupling) and use the
exact eigenenergies and eigenvectors as the input into the weak
coupling prescription,\cite{spo-rmp-80,accardi} as was recently
done in a dissipative double-dot system in
Ref.~\onlinecite{agu-prl-04}.

Rather than following this lengthy procedure, we used the standard
quantum optical damping kernel for a single harmonic oscillator in
the rotating wave approximation\cite{gardiner, carmichael} also
used in previous studies.\cite{arm-prb-02,bra-prb-03,modena}
Strictly speaking, this can be justified only in the case of weak
electromechanical coupling and small injection. Nevertheless, we
believe that the genuine non-equilibrium phenomena described later
on are captured qualitatively correctly even with this kernel
since the kernel mostly serves just as a ``convergence factor" to
stabilize the stationary solution. As will be seen below, the
sequential tunneling limit is extremely well captured within the
adopted approach. This is perhaps not too surprising since in that
limit the coherence between different dots is negligible. On the
other hand, the clear advantage of our choice of the damping
kernel is that it preserves charge conservation throughout the
whole circuit while this may not happen in general in the weak
coupling prescription (see Section \ref{equivalence}). Refinements
of the present approaches to deal with the above issues are in our
opinion a challenging task for the future modelling of NEMS. We
would like to point out that the above mentioned concerns about
additivity of the two baths apply also to the case of the one-dot
setup traditionally used for the description of the shuttling
phenomena \cite{nov-prl-03, fed-prl-04,nov-preprint-04} but the
problem stemming from the coherence present within the array is
absent there.

Bearing all these cautions in mind, we are ready to state the
generalized master equation \cite{arm-prb-02} for the $n$-resolved
density matrix of the system:
\begin{equation}\label{GME2}
\begin{split}
\dot{\hat{\rho}}^{(n)}_{00}&=-i[\hat{H}_{\rm
osc},\hat{\rho}_{00}^{(n)}] + \mathcal{L}_{\rm
damp}\hat{\rho}^{(n)}_{00} - \Gamma\hat{\rho}_{00}^{(n)}
+\Gamma\hat{\rho}_{RR}^{(n-1)} \,\\
\dot{\hat{\rho}}^{(n)}_{IJ}&=-i\langle I|[\hat{H}_{\rm el} +
\hat{H}_{\rm osc}+\hat{H}_{\rm el-osc},\hat{\rho}^{(n)}]|J\rangle
\\ &+ \mathcal{L}_{\rm damp}\hat{\rho}^{(n)}_{IJ} + \langle
I|\mathcal{L}_{\rm driv} \hat{\rho}^{(n)}|J\rangle\ \text{ for }
I,J=L,C,R\ .
\end{split}
\end{equation}
The commutator terms in \eqref{GME2} describe the coherent
evolution of the isolated device. The driving kernel $
\mathcal{L}_{\rm driv}$ is given just by substitution
$\hat{\sigma}\to\hat{\rho}$ in \eqref{driving}:
\begin{equation}\label{injection}
 \mathcal{L}_{\rm driv}\hat{\rho}=\Gamma\begin{pmatrix}
   \hat{\rho}_{00} & 0     & -\hat{\rho}_{LR}/2 \\
   0    & 0     & -\hat{\rho}_{CR}/2 \\
   -\hat{\rho}_{RL}/2 & -\hat{\rho}_{RC}/2  &  -\hat{\rho}_{RR}
 \end{pmatrix}.
\end{equation}
Finally, the damping kernel \cite{arm-prb-02} (acting as unity on
the electronic degrees of freedom) reads
\begin{equation}\label{damping}
\begin{split}
 \mathcal{L}_{\rm damp}\hat{\rho} &= -\frac{\gamma}{2}\bar{n}
 (\hat{a}\hat{a}^{\dagger}\hat{\rho} - 2 \hat{a}^{\dagger}\hat{\rho}\hat{a} +
 \hat{\rho}\hat{a}\hat{a}^{\dagger}) \\
 &-\frac{\gamma}{2}(\bar{n}+1)(\hat{a}^{\dagger}\hat{a}\hat{\rho}
  - 2 \hat{a}\hat{\rho}\hat{a}^{\dagger} + \hat{\rho}\hat{a}^{\dagger}\hat{a})
\end{split}
\end{equation}
where $\gamma$ is the damping rate and
$\bar{n}=n_B(\om_0)=(\exp(\om_0/k_B T)-1)^{-1}$ is the mean
occupation number of the oscillator at temperature $T$. This term
describes the effect of the environment on the oscillator,
consisting in mechanical damping and random quantum and thermal
excitation (Langevin force). The issue of the appropriate choice
of the damping kernel is, however, quite subtle in many respects
even in the case of a simple harmonic oscillator used here. There
is a well-known dilemma between the rotating wave approximation
form (conserving the positive definiteness of the resulting
density matrix) which we use in this work {\it versus} the
translationally invariant form (yielding correct equations of
motion for the mean coordinate and momentum) used
previously.\cite{nov-prl-03,nov-preprint-04} It is known that this
dilemma cannot be solved within the Markov approximation (without
relaxing the condition of approach to the canonical thermal
equilibrium state for asymptotic times; for a thorough discussion
of this issue see Ref.~\onlinecite{koh-jcp-97}).  We have carried
out a number of numerical checks, and have found out that in the
present case there are only minor differences in the obtained
results.   A practical advantage of the present choice is that it
leads to faster numerical convergence.

We can recast the GME \eqref{GME2} into a compact form
\begin{equation}\label{GME}
\begin{split}
\dot{\hat{\rho}}^{(n)}&=(\mathcal{L}-\mathcal{I}_{0R})\hat{\rho}^{(n)}
  +\mathcal{I}_{0R}\hat{\rho}^{(n-1)}\ ,\\
\dot{\hat{\rho}}&=\mathcal{L}\hat{\rho}\quad \text{ with }
\hat{\rho}=\sum_{n=0}^{\infty}\hat{\rho}^{(n)}\ \text{ and }
\hat{\rho}^{(-1)}\equiv 0
\end{split}
\end{equation}
where
$\mathcal{I}_{0R}\hat{\rho}=\Gamma\proj{0}{R}\hat{\rho}\proj{R}{0}$
(the symbol $\mathcal{I}_{0R}$ denotes the superoperator of the
particle current across the junction $0R$ between the right dot
and the right lead, for a discussion on superoperators see below).

The dynamics of the device described by the above generalized
master equation \eqref{GME} constitutes a {\em quantum Markov
process}.\cite{gardiner} The Liouvillean $\mathcal{L}$ determines
the evolution superoperator $\exp(\mathcal{L}t)$ which fully
characterizes the resulting quantum Markov process. It can be used
to calculate arbitrary multi-time correlation functions of any
{\em system operators}, i.e.\ operators acting as unity on the
Hilbert space of the reservoirs, by using the multi-time structure
of the quantum Markov process (often referred to as the {\em
quantum regression theorem}) --- for details see
Ref.~\onlinecite{gardiner}, Sec.\ 5.2 or
Ref.~\onlinecite{carmichael}, Sec.\ 3.2. Therefore, not only the
mean value of the stationary current within the array as in
Refs.~\onlinecite{arm-prb-02}, \onlinecite{bra-prb-03} can be
evaluated in this way, but also its higher order correlation
functions, in particular the current noise spectrum, become
accessible. The calculation can only be done for the junctions
{\em within} the array. For the outer junctions between the outer
dots and leads the quantum regression theorem cannot be applied
since the corresponding current operators involve the lead
electrons, thereby not being {\it system operators}. However, the
$n$-resolved form of the GME \eqref{GME} enables us to calculate
the current noise spectrum also for those junctions. Both methods
yield equivalent results as we will show later in Section
\ref{equivalence}.

\subsection{Notational details}

The linear operator $\mathcal{L}$ which acts on the density
operators, as specified by  \eqref{GME2}--\eqref{GME}, can be
handled (at least formally) as any other linear operator. We can
associate a matrix (infinite in our case) with it and perform
standard linear algebra operations. In order to avoid confusion
with ``normal" quantum mechanical operators acting in the
``normal" Hilbert space of the system, the vector space of
``normal" operators is called the Liouville space or the
superspace, and the Liouvillean and other linear operators acting
in the superspace are called superoperators (or supermatrices). In
the following, all superoperators will be denoted by calligraphic
symbols and the vectors of the superspace in the bra-ket notation
will be distinguished from the normal vectors in the Hilbert space
by double brackets, e.g.\ $\hat{V}\leftrightarrow
|v\rangle\!\rangle$ with $\hat{V}$ being a ``normal" quantum
mechanical operator.

If $\{|n\rangle\}_{n=1}^{\infty}$ is an orthonormal basis in the
Hilbert space of the system then all the projectors
$\{\proj{m}{n}\equiv|mn\rangle\!\rangle\}_{m,n=1}^{\infty}$ form
an orthonormal basis of the corresponding Liouville space with
respect to the scalar product $\langle\!\langle
a|b\rangle\!\rangle={\rm Tr}_{\rm sys}(\hat{A}^{\dag}\hat{B})$.
The matrix representation of superoperators follows analogously to
the normal Hilbert space case, i.e.\ $\mathcal{O}=\sum_{kl,
mn}|kl\rangle\!\rangle\langle\!\langle kl|\mathcal{O}
|mn\rangle\!\rangle\langle\!\langle mn|=\sum_{kl,
mn}|kl\rangle\!\rangle \,{\rm O}_{kl,mn}\langle\!\langle mn|$.
There is a unique mapping between matrices representing the
operators in the Hilbert space and the vectors in the Liouville
space. The operator $\hat{O}=\sum_{k,l}|k\rangle O_{kl}\langle l|$
represented by the matrix $\left(\begin{smallmatrix} O_{11} &
O_{12} & \dots \\O_{21} & O_{22} & \dots \\ \vdots & \vdots &
\ddots
\end{smallmatrix}\right)$ corresponds to the vector
$|o\rangle\!\rangle=\sum_{kl}O_{kl}|kl\rangle\!\rangle$
represented by the column vector
$\mathbf{o}=(O_{11},O_{12},O_{13},\dots,O_{21},O_{22},O_{23},\dots)^T$
(the exact ordering depends on the chosen ordering of the double
indices $kl$). Therefore, we will in the following use the two
representations interchangeably.

\section{Noise calculation}
\label{theory}

\subsection{Definition and properties of the current noise spectrum}
\label{definitions}

In this subsection we define the current noise spectra for
different junctions present in our model and analyze several of
their properties. First, we find the current operators across
different junctions. From the equations of motion for the
operators of the occupation of the respective dots
$\hat{n}_J=\proj{J}{J},\,J=0,L,C,R$ reading
\begin{equation}\label{charge_cons}
    e\frac{d}{dt}\,\hat{n}_{J}=-ie[\hat{n}_{J},\hat{H}]=\hat{I}_{J+}-\hat{I}_{J-}
\end{equation}
we identify the corresponding charge current operators (electronic
charge is $e<0$; electrons flow from left to right)
\begin{subequations}\label{currents}
\begin{align}
\begin{split}
    \hat{I}_{0-}&\equiv\hat{I}_{L+}\equiv\hat{I}_{L0}= -e\frac{d}{dt}\,\hat{N}_L(t)\\
        &= ie\sum_{k} V_{kL}\Bigl(\hat{c}_{kL}^{\dag}\proj{0}{L}
       - \proj{L}{0}\hat{c}_{kL}\Bigr)\ ,
\end{split}\label{current0L}\\
  \hat{I}_{L-}&\equiv\hat{I}_{C+}\equiv\hat{I}_{CL}=
  ie t_L(\hat{x})\big(\proj{L}{C}-\proj{C}{L}\big)
    \ ,\label{currentLC}\\
    \hat{I}_{C-}&\equiv\hat{I}_{R+}\equiv\hat{I}_{RC}=
    ie t_R(\hat{x})\big(\proj{C}{R}-\proj{R}{C}\big)
    \ ,\label{currentCR} \\
\begin{split}
    \hat{I}_{R-}&\equiv\hat{I}_{0+}\equiv\hat{I}_{0R}= e\frac{d}{dt}\,\hat{N}_R(t)\\
    &=ie\sum_{k} V_{kR}\Bigl(\proj{R}{0}\hat{c}_{kR}
    -\hat{c}_{kR}^{\dag}\proj{0}{R}\Bigr)
\end{split} \label{currentR0}
\end{align}
\end{subequations}
with $\hat{N}_L=\sum_k\hat{c}^{\dag}_{kL}\hat{c}_{kL},\,
\hat{N}_R=\sum_k\hat{c}^{\dag}_{kR}\hat{c}_{kR}$ being the
operators of the number of particles in the left and right lead,
respectively.

We next define different current-current correlation functions
($a,b=L0,CL,RC,0R$)
\begin{equation}\label{noise_def}
\begin{split}
C_{ab}(\tau)&=
\lim_{t\to\infty}\Big[\frac{1}{2}\langle\{\hat{I}_a(t+\tau),\hat{I}_b(t)\}\rangle
-\langle \hat{I}_a(t+\tau )\rangle\langle \hat{I}_b(t)\rangle\Big]\\
&=\lim_{t\to\infty}\frac{1}{2}\langle\{\Delta\hat{I}_a(t+\tau),\Delta\hat{I}_b(t)\}\rangle
\ , \\ &\text{ with }
\Delta\hat{I}_a(t)=\hat{I}_a(t)-\langle\hat{I}_a(t)\rangle
\end{split}
\end{equation}
which in the stationary limit are functions of $\tau$ only. We
also note the property $C_{ab}(-\tau)=C_{ba}(\tau)$. The current
noise spectrum is\footnote{There are different conventions used in
the literature. Here we define the spectrum as the Fourier
transform of the correlation function without a prefactor of $2$
used in, e.g.\ in Ref.~\onlinecite{bla-phr-00}. The Fano factor is
then given as $F(0)=\tfrac{S(0)}{eI}$.}
\begin{equation}\label{noise_spect}
    S_{ab}(\om) = \int_{-\infty}^{\infty}d\tau C_{ab}(\tau)e^{i\omega\tau}
    \ .
\end{equation}
The diagonal elements $S_{aa}(\om)$ of the noise matrix are
non-negative as can be shown by using the Lehmann representation.

In general, for an arbitrary frequency the noise depends on the
position where the current is measured. However, in the limit
$\om\to 0$ charge conservation implies that the noise becomes
independent of the measurement position along the circuit, i.e.\
$S_{aa}(0)=S_{bb}(0)=S_{ab}(0)=S_{ba}(0),\,a\neq b$ and it also
equals the shot noise component of the spectrum measured in the
leads. This statement is proven by considering current correlation
functions for two adjacent junctions $J+,\,J-$.\footnote{A similar
proof has been suggested to us by G. Kie{\ss}lich (private
communication).} The charge conservation condition
\eqref{charge_cons} gives
\begin{equation}
\begin{split}
C_{J+J+}(\tau)&=\frac{1}{2}\langle\{\Delta\hat{I}_{J+}(\tau),\Delta\hat{I}_{J+}\}\rangle
=\frac{1}{2}\langle\{\Delta\hat{I}_{J-}(\tau),\Delta\hat{I}_{J+}\}\rangle\\
&+\frac{1}{2}\frac{d}{d\tau}\langle\{e\Delta\hat{n}_{J}(\tau),\Delta\hat{I}_{J+}\}\rangle\\
&=C_{J-J+}(\tau)+\frac{1}{2}\frac{d}{d\tau}\langle
\{e\Delta\hat{n}_{J}(\tau),\Delta\hat{I}_{J+}\}\rangle
\end{split}
\end{equation}
which implies $S_{J+J+}(0)=S_{J-J+}(0)$. The relation
$C_{ab}(-\tau)=C_{ba}(\tau)$ yields $S_{J-J+}(-\om)=S_{J+J-}(\om)$
and by using the charge conservation again we can finally
establish $S_{J+J-}(0)=S_{J-J-}(0)$. Altogether we find that the
zero-frequency noise is the same for any combination of the
junctions, i.e.\ $S_{ab}(0)=S(0)\geq 0$ for any $a,\,b$ (not
necessarily adjacent; this generalization is straightforward).

The current operators $\hat{I}_{CL},\,\hat{I}_{RC}$
\eqref{currentLC}, \eqref{currentCR} between the dots are
obviously system operators in the sense that they operate as unity
on the degrees of freedom of the leads and the heat bath.
Therefore, we can use the formalism of quantum Markov processes to
evaluate correlation functions involving these operators using the
quantum regression theorem --- this will be done in subsection
\ref{qrth}. This is not the case for the operators of current
between the outer dots and leads $\hat{I}_{L0},\,\hat{I}_{0R}$
given by \eqref{current0L}, \eqref{currentR0}. However, the noise
spectra across these two junctions can still be calculated using
the $n$-resolved form of the GME \eqref{GME} with the help of the
following identity for the zero-frequency current noise (for the
junction $0R$, the case $L0$ is analogous)
\begin{equation}\label{mcd_formula}
\begin{split}
    &\frac{d}{dt}\Big(\langle\hat{Q}_R^2(t)\rangle-\langle\hat{Q}_R(t)\rangle^2\Big)
    \Big|_{t\to\infty}=\int_{-\infty}^{\infty}d\tau C_{0R,0R}(\tau)=S_{0R,0R}(0)\\
    &\text{ with }
    \hat{Q}_R(t)=e\hat{N}_R(t)-e\hat{N}_R(0)=\int_0^t dt'\hat{I}_{0R}(t')
    \ .
\end{split}
\end{equation}
This identity suggests the interpretation of the zero-frequency
current noise as the ``charge diffusion
coefficient"\cite{cam-preprint-04} and will be used in subsection
\ref{mcd} for an alternative evaluation of the zero-frequency
current noise. The equivalence of the two approaches is shown
explicitly in subsection \ref{equivalence}.

We finally comment on the physical relevance of the noise spectra
calculated in this paper. Since the zero-frequency noise is
position-independent the noise calculated for the junctions within
the system should also be measured in the leads. However, in
practice there is always the important $1/f$ contribution to the
noise which actually dominates experiments for very low
frequencies and which is not accounted for in our model.
Therefore, as mentioned in Ref.~\onlinecite{dav-prb-92}, the
measurements of the shot noise must be performed at non-zero
frequencies of the order of 1 kHz where the $1/f$ noise component
becomes insignificant. However, the shot noise measured in this
way is still appropriately described by the zero-frequency current
noise calculations since its typical frequency scale is of the
order of 1 THz.

\subsection{Quantum regression theorem (QRT)}
\label{qrth}

With QRT it is possible to calculate the current noise {\em
within} the system (i.e.\ for $\hat{I}_{CL},\,\hat{I}_{RC}$). For
$\tau\ge 0$ QRT gives (cf.\ Ref.~\onlinecite{gardiner}, Sec.\ 5.2)
\begin{equation}\label{qrt}
C_{ab}(\tau)=\frac{1}{2}\mathrm{Tr}_{\rm sys} (\hat{I}_a
\exp(\mathcal{L}\tau)\{\hat{I}_b,\hat{\rho}^{\rm stat}\})-I^2
\end{equation}
for $a,b=CL,RC$, where
$I=\lim_{t\to\infty}\langle\hat{I}_a(t)\rangle=\mathrm{Tr}_{\rm
sys}(\hat{I}_a\hat{\rho}^{\rm stat})$ is the stationary current
(constant throughout the circuit). In case $\tau<0$ we use the
symmetry property $C_{ab}(-\tau)=C_{ba}(\tau)$. Now, let us
evaluate the spectrum
\begin{equation}
\begin{split}
    S_{ab}(\om) &= \int_{-\infty}^{\infty}d\tau C_{ab}(\tau)e^{i\omega\tau}\\
        & =\int_0^{\infty}d\tau C_{ab}(\tau)e^{i\omega\tau}
          +\int_0^{\infty}d\tau C_{ba}(\tau)e^{-i\omega\tau} \ .
\end{split}
\end{equation}
We consider in detail the first term denoted $S_{ab}^+(\om)$, the
second one ($S_{ba}^-(\om)$) follows analogously. Introducing a
convergence factor $\om\to\om + i0$ we get
\begin{equation}\label{spect}
S_{ab}^+(\om)= \frac{1}{2}\mathrm{Tr}_{\rm sys} (\hat{I}_a
(-i\om-\mathcal{L})^{-1}\{\hat{I}_b,\hat{\rho}^{\rm stat}\})
+\frac{1}{i\om}I^2.
\end{equation}

Since we are interested in the limit $\om\to 0$ in the end we have
to handle somehow the singularities associated with the resolvent
$\mathcal{G}(-i\om)=(-i\om-\mathcal{L})^{-1}$ and the second term
in \eqref{spect} in that limit. The problem with the inverse of
$\mathcal{L}$ is the existence of the unique null vector
$|0\rangle\!\rangle$ which is proportional to the stationary
density matrix because $\mathcal{L}\hat{\rho}^{\rm stat}=0$. There
exists a corresponding left eigenvector belonging to the zero
eigenvalue of $\mathcal{L}$ denoted by
$\langle\!\langle\widetilde{0}|$ which is not just the hermitian
conjugate of $|0\rangle\!\rangle$ (i.e.\
$\langle\!\langle\widetilde{0}|\neq|0\rangle\!\rangle^{\dag}$)
because $\mathcal{L}$ is non-hermitian. However, since
$\mathrm{Tr}_{\rm sys}(\mathcal{L}\hat{A})=0$ for any system
operator $\hat{A}$ we deduce that
$\langle\!\langle\widetilde{0}|\leftrightarrow\hat{1}$, i.e.\
$\langle\!\langle\widetilde{0}|\mathcal{L}|a\rangle\!\rangle\equiv\mathrm{Tr}_{\rm
sys}(\hat{1}\,\mathcal{L}\hat{A})=0$.

Thus, we have $|0\rangle\!\rangle \leftrightarrow \hat{\rho}^{\rm
stat},\, \langle\!\langle\widetilde{0}|\leftrightarrow \hat{1}$
with $\langle\!\langle\widetilde{0}|0\rangle\!\rangle=1$ allowing
us to define the projectors
$\mathcal{P}=\mathcal{P}^2=|0\rangle\!\rangle\!\langle\!\langle\widetilde{0}|,
\,\mathcal{Q}=1-\mathcal{P}$. Using these projectors and the
relations $\mathcal{P}\mathcal{L}=\mathcal{L}\mathcal{P}=0,\,
\mathcal{L}=\mathcal{Q}\mathcal{L}\mathcal{Q}$ the resolvent can
be expressed as
\begin{equation}\label{resolvent}
\begin{split}
    \mathcal{G}(-i\om)&=(-i\om-\mathcal{L})^{-1}=(-i\om\mathcal{P}-i\om\mathcal{Q}
    -\mathcal{Q}\mathcal{L}\mathcal{Q})^{-1}\\
    &=-\frac{1}{i\om}\mathcal{P}
    -\mathcal{Q}\frac{1}{i\om+\mathcal{L}}\mathcal{Q}\\
    &\approx -\frac{1}{i\om}\mathcal{P}-\mathcal{Q}\mathcal{L}^{-1}\mathcal{Q}
    =-\frac{1}{i\om}\mathcal{P}-\mathcal{R}\
    \text{ for small}\ \om
\end{split}
\end{equation}
where we have defined the pseudoinverse of the Liouvillean
$\mathcal{R}\equiv\mathcal{Q}\mathcal{L}^{-1}\mathcal{Q}$.
Substituting the term $-\tfrac{i\mathcal{P}}{\om}$ in the first
term of \eqref{spect} gives
\begin{equation}
\begin{split}
    &-\frac{1}{2i\om}\mathrm{Tr}_{\rm sys} (\hat{I}_a|0\rangle\!\rangle
    \!\langle\!\langle\widetilde{0}|\{\hat{I}_b,\hat{\rho}^{\rm  stat}\})\\
    &= -\frac{1}{2i\om} \mathrm{Tr}_{\rm sys}(\hat{I}_a \hat{\rho}^{\rm stat})
    \mathrm{Tr}_{\rm sys}\bigl(\{\hat{I}_b,\hat{\rho}^{\rm stat}\}\bigr)
    = -\frac{1}{i\om} I^2,
\end{split}
\end{equation}
which cancels the last term of \eqref{spect}. Applying the same
procedure to $S_{ba}^-(0)$ we find
\begin{equation}\label{noise}
\begin{split}
 &S_{ab}(0)=S_{ab}^+(0)+S_{ba}^-(0)\\
 &= -\frac{1}{2}\big(\mathrm{Tr}_{\rm sys} (\hat{I}_a \mathcal{R}
 \{\hat{I}_b,\hat{\rho}^{\rm stat}\})+\mathrm{Tr}_{\rm sys} (\hat{I}_b \mathcal{R}
 \{\hat{I}_a,\hat{\rho}^{\rm stat}\})\big).
\end{split}
\end{equation}
If we introduce the superoperators of (particle) current
$\mathcal{I}_{CL},\,\mathcal{I}_{RC}$ defined by their action on
the system density matrix as follows
$e\mathcal{I}_a\hat{\rho}=\tfrac{1}{2}\{\hat{I}_a,\hat{\rho}\},\,a=CL,RC$
with the property $I=e\mathrm{Tr}_{\rm sys}\mathcal{I}_a
\hat{\rho}^{\rm stat}= e\langle\!\langle\tilde{0}|\mathcal{I}_{a}
|0\rangle\!\rangle$ we can rewrite the above equation in a compact
form
\begin{equation}\label{noiseLC}
    S_{ab}(0)=-e^2\langle\!\langle\tilde{0}|\mathcal{I}_{a}\mathcal{R}\mathcal{I}_{b}
    +\mathcal{I}_{b}\mathcal{R}\mathcal{I}_{a}|0\rangle\!\rangle\
    \ a,b=CL,RC\ .
\end{equation}
This equation constitutes the main formal result of this
subsection and forms the basis for further formal manipulations
and eventually the numerical treatment.

\subsection{Counting variable approach --- evaluation of the charge diffusion coefficient}
\label{mcd}

Using the $n$-resolved form of the GME \eqref{GME} we could in
principle find the full counting statistics (FCS) of the charge
transfer through the junction between the right dot and the right
lead, i.e.\ the probabilities $P_n(t)$ that $n$ electrons tunneled
into the right lead across the junction by time $t$ given by
$P_n(t)=\mathrm{Tr_{sys}}\hat{\rho}^{(n)}(t)$. Here, we are only
interested in the evaluation of the zero-frequency noise for which
we just need the mean and the mean square charge tunneled into the
right lead by time $t$ given by
$\langle\hat{Q}_R(t)\rangle=e\sum_n
nP_n(t),\,\langle\hat{Q}_R^2(t)\rangle=e^2\sum_n n^2P_n(t)$. Using
the definition of the current \eqref{currentR0} and the identity
\eqref{mcd_formula} we find the stationary mean current and the
zero-frequency current noise:\cite{ela-pla-02}
\begin{align}
    &I_{0R} = e \frac{d}{dt}\sum_nn P_n(t)\Big|_{t\to\infty}
    = e \sum_n n \dot{P}_n(t)\Big|_{t\to\infty},\label{current}\\
    &S_{0R,0R}(0) = e^2\frac{d}{dt}\bigg[\sum_nn^2P_n(t)
    -\Big(\sum_nnP_n(t)\Big)^2\bigg]\bigg|_{t\to\infty}\notag\\
    &=e^2\bigg[\sum_nn^2\dot{P}_n(t)-2\Big(\sum_nnP_n(t)\Big)
    \Big(\sum_nn\dot{P}_n(t)\Big)\bigg]\bigg|_{t\to\infty}.
    \label{macdonald}
\end{align}

We evaluate $\dot{P}_n(t)$ from Eq.~\eqref{GME} and find
\begin{align}
  &\dot{P}_n(t) = \mathrm{Tr_{sys}}\big[\mathcal{I}_{0R}
  \big(\hat{\rho}^{(n-1)}(t)-\hat{\rho}^{(n)}(t)\big)\big]\\
  \intertext{and consequently}
  &\sum_n \dot{P}_n(t) = 0 \ ,\\
  &\sum_n n \dot{P}_n(t) = \mathrm{Tr_{sys}}
  \Big(\mathcal{I}_{0R}\sum_n\hat{\rho}^{(n)}(t)\Big)
  =\mathrm{Tr_{sys}}\big(\mathcal{I}_{0R}\hat{\rho}(t)\big)\ ,\\
  &\sum_n n^2 \dot{P}_n(t) = \mathrm{Tr_{sys}}\Big[\mathcal{I}_{0R}
  \Big(2\sum_n n \hat{\rho}^{(n)}(t) + \hat{\rho}(t)\Big)\Big] \ ,
\end{align}
where according to the definition $\sum_n
\hat{\rho}^{(n)}(t)=\hat{\rho}(t)$. Now, we employ an
operator-valued generalization of the standard generating function
technique to calculate $\sum_n n \hat{\rho}^{(n)}(t)$. We
introduce the object $\hat{F}(t;z)=\sum_n \hat{\rho}^{(n)}(t)z^n$
which has the properties
$\hat{F}(t;1)=\hat{\rho}(t),\,\tfrac{\partial}{\partial
z}\hat{F}(t;z)|_{z=1}=\sum_n n \hat{\rho}^{(n)}(t)$ and satisfies
the equation of motion
\begin{equation}\label{EOM}
    \frac{\partial}{\partial t}\hat{F}(t;z)=\big(\mathcal{L}
    +(z-1)\mathcal{I}_{0R}\big)\hat{F}(t;z)\ .
\end{equation}
Using the generating function the current noise formula
\eqref{macdonald} can be rewritten as
\begin{equation}\label{noisegen}
\begin{split}
&S_{0R,0R}(0)=e^2\bigg(\mathrm{Tr_{sys}}\Big[\mathcal{I}_{0R}\Big(2\frac{\partial}
{\partial z}\hat{F}(t;z)\Big|_{z=1}+ \hat{F}(t;1)\Big)\Big]\\
&-2\mathrm{Tr_{sys}}\Big(\mathcal{I}_{0R}\hat{F}(t;1)\Big)
\mathrm{Tr_{sys}}\Big(\frac{\partial}{\partial z}
\hat{F}(t;z)\Big|_{z=1}\Big)\bigg)\bigg|_{t\to\infty}.
\end{split}
\end{equation}
The equation of motion for $\hat{F}(t;z)$ \eqref{EOM} can be
solved via the Laplace transform
$\tilde{\hat{F}}(s;z)=\int_0^{\infty}dt e^{-st}\hat{F}(t;z)$
giving
\begin{equation}
    \big(s-\mathcal{L}-(z-1)\mathcal{I}_{0R}\big)\tilde{\hat{F}}(s;z)=
     \sum_n \hat{\rho}^{(n)}(0)z^n\ ,
\end{equation}
with $\hat{\rho}^{(n)}(0)$ being the initial conditions. Recalling
the definition of the resolvent
$\mathcal{G}(s)=(s-\mathcal{L})^{-1}$ of the Liouvillean we arrive
at
\begin{align}
    \tilde{\hat{F}}(s;1)&=\mathcal{G}(s)\hat{\rho}(0)\\
    \frac{\partial}{\partial z}\left.\tilde{\hat{F}}(s;z)\right|_{z=1}&=
    \mathcal{G}(s)\mathcal{I}_{0R}\mathcal{G}(s)\hat{\rho}(0)
    +\mathcal{G}(s)\sum_n n \hat{\rho}^{(n)}(0)\ .
\end{align}
Because the large-$t$ behavior of $\hat{F}(t;z)$ is related to the
small-$s$ behavior of $\tilde{\hat{F}}(s;z)$ we study the
asymptotics of the above expressions as $s\to 0+$. This is
entirely determined by the resolvent $\mathcal{G}(s)$ in the
small-$s$ limit. We can use the results from the previous
subsection and substitute $-i\om\to s$ to get the leading
asymptotics of $\mathcal{G}(s)$ for $s\to 0+$. Thus, we obtain
\begin{align}
    \tilde{\hat{F}}(s;1)&\approx\frac{\mathcal{P}}{s}\hat{\rho}(0)
    =\frac{1}{s}\hat{\rho}^{\rm stat}\\
    \frac{\partial}{\partial z}\left.\tilde{\hat{F}}(s;z)\right|_{z=1}
    &\approx \frac{1}{s^2}\mathcal{P}\mathcal{I}_{0R}\mathcal{P}\hat{\rho}(0)
    -\frac{1}{s}\Big[\mathcal{P}\mathcal{I}_{0R}\mathcal{R}\hat{\rho}(0)\notag\\
    &+\mathcal{R}\mathcal{I}_{0R}\mathcal{P}\hat{\rho}(0)
    -\mathcal{P}\sum_n n \hat{\rho}^{(n)}(0)\Big]\ .
\end{align}
In the time domain this gives
\begin{align}
    \hat{F}(t;1)|_{t \to\infty}&\approx\hat{\rho}^{\rm stat}\\
    \frac{\partial}{\partial z}\left.\hat{F}(t;z)\right|_{z=1,t\to\infty}
    &\approx \hat{\rho}^{\rm stat}\Big(\frac{I}{e}\,t+C^{\rm init}\Big)
    -\mathcal{R}\mathcal{I}_{0R}\hat{\rho}^{\rm stat}\ ,
\end{align}
where $C^{\rm init}=\mathrm{Tr_{sys}}\big(\sum_n n
\hat{\rho}^{(n)}(0)-\mathcal{I}_{0R}\mathcal{R}\hat{\rho}(0)\big)$
is an initial conditions dependent constant and the stationary
current is given by
$I=e\mathrm{Tr_{sys}}(\mathcal{I}_{0R}\hat{\rho}^{\rm stat})$. The
corrections to the large time asymptotic behavior are
exponentially small --- the approach to the stationary state in a
Markovian system is exponential. In particular, it is important
that there is no $\tfrac{1}{t}$ correction to $\hat{F}(t;1)|_{t
\to\infty}$ (which would correspond to a $\ln s$-like divergence
in the resolvent as $s\to 0+$) since it would combine with the
linearly in $t$ divergent term in $\tfrac{\partial}{\partial
z}\left.\hat{F}(t;z)\right|_{z=1,t\to\infty}$ to yield a finite
term in \eqref{noisegen}. We substitute the above asymptotic
formulas into Eq.~\eqref{noisegen}, use the definition of the
stationary current and the identities
$\mathrm{Tr_{sys}}\hat{\rho}^{\rm
stat}=1,\,\mathrm{Tr_{sys}}\mathcal{R}\bullet=0$  to get the final
result for the zero-frequency current noise at the $0R$ junction,
\begin{equation}\label{noise0R}
\begin{split}
   S_{0R,0R}(0)&=eI - 2e^2\mathrm{Tr_{sys}}\big(\mathcal{I}_{0R}\mathcal{R}
   \mathcal{I}_{0R}\hat{\rho}^{\rm stat}\big)\\
   &= e^2\langle\!\langle\tilde{0}|\mathcal{I}_{0R}
   -2\mathcal{I}_{0R}\mathcal{R}\mathcal{I}_{0R}|0\rangle\!\rangle\ .
\end{split}
\end{equation}
In the algebra leading to \eqref{noise0R} the linearly divergent
terms in $t$ and the initial condition terms cancel identically so
that we are left with a regular, initial-condition-independent
expression as expected and necessary. Similarly, for the $L0$
junction one finds
\begin{equation}\label{noiseL0}
\begin{split}
   S_{L0,L0}(0)&=eI - 2e^2\mathrm{Tr_{sys}}\big(\mathcal{I}_{L0}\mathcal{R}
   \mathcal{I}_{L0}\hat{\rho}^{\rm stat}\big)\\
   &= e^2\langle\!\langle\tilde{0}|\mathcal{I}_{L0}
   -2\mathcal{I}_{L0}\mathcal{R}\mathcal{I}_{L0}|0\rangle\!\rangle
\end{split}
\end{equation}
with
$\mathcal{I}_{L0}\hat{\rho}=\Gamma\proj{L}{0}\hat{\rho}\proj{0}{L}$.

\subsection{Equivalence of different approaches}
\label{equivalence}

We show the equality between the expressions \eqref{noiseLC} and
\eqref{noise0R}, \eqref{noiseL0}. Both formulas contain the same
basic building block consisting of terms of the type
$\mathcal{I}_{a}\mathcal{R}\mathcal{I}_{b}$. However, there is an
obvious difference:  the presence of the so-called
self-correlation or Schottky term (proportional to the mean
current) in formulas \eqref{noise0R}, \eqref{noiseL0}. Yet, they
give the same value for the zero-frequency noise in the end as we
now proceed to show.

The independence of the zero-frequency noise from the position
along the circuit has been shown quite generally in subsection
\ref{definitions} using the charge conservation. Thus, the only
task now is to find the corresponding expression for the charge
conservation within the superoperator language. Following the
purely stochastic analogy \cite{her-prb-93} we find that the
charge conservation condition \eqref{charge_cons} is expressed in
terms of superoperators by the following equation
\begin{equation}\label{ch-cons}
    [\mathcal{N}_J,\mathcal{L}]=\mathcal{I}_{J+}-\mathcal{I}_{J-}
\end{equation}
with the superoperators of occupation of the ``site" $J,\
J=0,L,C,R$ being given by
$\mathcal{N}_J\hat{\rho}=\tfrac{1}{2}\{\proj{J}{J},\hat{\rho}\}$,
the current superoperators $\mathcal{I}_{a}$ were defined
previously and the convention for $J\pm$ is the same as in
Eqs.~\eqref{currents}. The above relation follows from the
definitions of the respective quantities and equations
\eqref{GME2}--\eqref{charge_cons}.

Since the heat bath does not couple directly to the electronic
degrees of freedom its degrees of freedom do not enter explicitly
the current and occupation operators, cf.\ \eqref{charge_cons} and
\eqref{currents}, and are therefore absent from the corresponding
superoperators. We believe that this property should be reflected
in the identity $[\mathcal{N}_J,\mathcal{L}_{\rm damp}]=0$ for any
choice of the damping kernel. Obviously, this condition is
fulfilled for our choice of the damping kernel \eqref{damping}.
However, for the generic weak coupling
prescription\cite{spo-rmp-80,accardi} for the damping kernel the
above identity may not be satisfied which would break the charge
conservation.\footnote{We have made preliminary tests of this
issue in a simpler model of the dissipative double-dot system
studied in Ref.~\onlinecite{agu-prl-04}. We have found out that
the zero-frequency noise for the inner junction between the dots
differs from the spectra of the outer junctions which are the
same. This difference can be understood from the relation
$[\mathcal{N}_{L,R},\mathcal{L}_{\rm damp}]\neq 0$ implied by that
model (cf.\ Ref.~\onlinecite{agu-prl-04}, Eqs.~(3), (4)). On the
other hand, the equivalence of the outer junctions is a
consequence of $[\mathcal{N}_L+\mathcal{N}_R,\mathcal{L}_{\rm
damp}]=0$.} This raises the possibility that there is another
problem with the Markovian weak damping prescription analogous to
the translational invariance issue threatening the charge
conservation for damped NEMS involving coherent charge transfer
(such as our quantum dot array). This issue deserves further
investigation.

The charge conservation relation \eqref{ch-cons} is used to prove
the position independence of the mean current
$I=e\langle\!\langle\tilde{0}|\mathcal{I}_{a}|0\rangle\!\rangle$
and the zero-frequency noise $S_{ab}(0)$ for any $a,b$. The mean
current conservation follows from
\begin{equation}
\begin{split}
    I&=e\langle\!\langle\tilde{0}|\mathcal{I}_{J+}|0\rangle\!\rangle
    =e\langle\!\langle\tilde{0}|\mathcal{I}_{J-}|0\rangle\!\rangle
    +e\langle\!\langle\tilde{0}|[\mathcal{N}_J,\mathcal{L}]|0\rangle\!\rangle\\
    &=e\langle\!\langle\tilde{0}|\mathcal{I}_{J-}|0\rangle\!\rangle
    \ \text{ due to }\mathcal{L}|0\rangle\!\rangle=0,\,
    \langle\!\langle\tilde{0}|\mathcal{L}=0\ .
\end{split}
\end{equation}
Analogously, we prove the equivalence, for example, between
$S_{0R,0R}(0)$ \eqref{noise0R} and $S_{RC,RC}(0)$ \eqref{noiseLC}.
Substituting \eqref{ch-cons} for $J=R$ into the expression
\eqref{noiseLC} for $S_{RC,RC}(0)$ we get in the first step
\begin{equation}
\begin{split}
     &S_{RC,RC}(0) = -2e^2\langle\!\langle\tilde{0}|\mathcal{I}_{RC}\mathcal{R}
     \mathcal{I}_{RC}|0\rangle\!\rangle\\
     &=e^2\langle\!\langle\tilde{0}|[\mathcal{I}_{RC},
     \mathcal{N}_R]|0\rangle\!\rangle
     -e^2\langle\!\langle\tilde{0}|\mathcal{I}_{0R}\mathcal{R}
     \mathcal{I}_{RC}+\mathcal{I}_{RC}\mathcal{R}
     \mathcal{I}_{0R}|0\rangle\!\rangle\\
     &=-e^2\langle\!\langle\tilde{0}|\mathcal{I}_{0R}\mathcal{R}
     \mathcal{I}_{RC}+\mathcal{I}_{RC}\mathcal{R}
     \mathcal{I}_{0R}|0\rangle\!\rangle\\
     &\equiv S_{RC,0R}(0)=S_{0R,RC}(0)
\end{split}
\end{equation}
bearing in mind $\mathcal{L}\mathcal{R}=\mathcal{R}\mathcal{L}=
\mathcal{Q}=1-|0\rangle\!\rangle\langle\!\langle\widetilde{0}|$
and finding $e[\mathcal{I}_{RC},\mathcal{N}_{R}]\hat{\rho}=
\tfrac{1}{4}[[\hat{I}_{RC},\proj{R}{R}],\hat{\rho}]$ which yields
zero when traced over. We proceed similarly in the second step and
obtain
\begin{equation}
    S_{0R,0R}(0) = -2e^2\langle\!\langle\tilde{0}|\mathcal{I}_{0R}\mathcal{R}
     \mathcal{I}_{0R}|0\rangle\!\rangle+e^2\langle\!\langle\tilde{0}|[\mathcal{I}_{0R},
     \mathcal{N}_R]|0\rangle\!\rangle\ .
\end{equation}
The second term can be evaluated as
$[\mathcal{I}_{0R},\mathcal{N}_R]=[\mathcal{N}_{0},\mathcal{I}_{0R}]=\mathcal{I}_{0R}$
recovering finally the expression \eqref{noise0R} for
$S_{0R,0R}(0)$.

By extending the argument to other combinations of the junctions
we can summarize the formulas for the zero-frequency noise
$S(0)=S_{I+,J+}(0)$ for any $I,J=0,L,C,R$ in the compact form as
(compare with the analogous expression for the purely stochastic
case in Ref.~\onlinecite{her-prb-93}, Eq. (26))
\begin{equation}\label{noise-general}
\begin{split}
    S(0)&=-e^2\langle\!\langle\tilde{0}|\mathcal{I}_{I+}\mathcal{R}
     \mathcal{I}_{J+}+\mathcal{I}_{J+}\mathcal{R}
     \mathcal{I}_{I+}|0\rangle\!\rangle\\
     &+\delta_{IJ}e^2\langle\!\langle\tilde{0}|[\mathcal{N}_{I},
     \mathcal{I}_{J+}]|0\rangle\!\rangle\ \text{ for any }I,J\ .
\end{split}
\end{equation}
This equation merges the two approaches into a single picture
unifying both the pure quantum mechanical and pure classical
stochastic formalisms. It has a quantum-mechanical-like form of a
``mean value" of the pseudoinverse of the Liouvillean
symmetrically flanked by two current superoperators corrected with
the classical-like self-correlation term. The self-correlation
term is only effective for the diagonal elements of the
current-current correlation matrix and, moreover, is non-zero just
for the outer junctions where it contributes by the mean current.

\subsection{Notes on numerical evaluation}
\label{numerics}

From the results obtained thus far we see that the evaluation of
the noise involves two steps. At the first step we find the
stationary state $\hat{\rho}^{\rm
stat}=\lim_{t\to\infty}\exp(\mathcal{L}t)\hat{\rho}_0$ independent
of the initial condition $\hat{\rho}_0$ and equivalently given by
the equation
\begin{equation}\label{stat}
 \mathcal{L}\hat{\rho}^{\rm stat}=0\ ,\ \mathrm{Tr}_{\rm sys}\hat{\rho}^{\rm stat}=1\ .
\end{equation}
Having found $\hat{\rho}^{\rm stat}$ we can fully characterize all
one-time quantities pertaining to the system such as occupations
of the different dots, mean current, Wigner functions of the
oscillator in different charge states etc.

To evaluate the noise (second step) we have to find the
pseudoinverse of the Liouvillean
$\mathcal{R}=\mathcal{Q}\mathcal{L}^{-1}\mathcal{Q}$. In practice,
we actually do not have to evaluate the whole pseudoinverse but we
fix a given combination of junctions and evaluate the auxiliary
quantities
$\hat{\Sigma}_a=e\mathcal{R}\mathcal{I}_a\hat{\rho}^{\rm stat}$
determined by the equation
\begin{equation}\label{gmres}
 \mathcal{L}\hat{\Sigma}_a =e\mathcal{I}_a\hat{\rho}^{\rm stat}
 - I \hat{\rho}^{\rm stat}\ ,\ \mathrm{Tr}_{\rm sys}\hat{\Sigma}_a=0\ .
\end{equation}
Eq.\ \eqref{gmres} has a solution since the right hand side lies
in the range of $\mathcal{L}$ (the trace of the right hand side is
zero) and the freedom of adding any multiple of the null vector to
a particular solution is fixed uniquely by the trace condition
$\mathrm{Tr}_{\rm sys}\Sigma_a=0$. Of course, this is equivalent
to the uniqueness and regularity of the standalone pseudoinverse
$\mathcal{R}$. Moreover, $\mathcal{R}$ preserves hermiticity so
that the quantities $\hat{\Sigma}_a$ are Hermitian as they should
be to give a real zero-frequency noise. This follows from the
property $(\mathcal{L}\hat{A})^{\dag}=\mathcal{L}\hat{A}$ for any
Hermitian $\hat{A}$ and the trace-fixing condition
$\mathrm{Tr}_{\rm sys}\hat{\Sigma}_a=0$ of Eq.\ \eqref{gmres}.

Equations \eqref{stat} and \eqref{gmres} form the starting point
for the numerical implementation of the noise calculation. After
the truncation of the oscillator Hilbert space to the $N$ lowest
energy states the size of the supermatrix $\mathcal{L}$
becomes\footnote{Remember that the block elements
$\rho_{0I},\,\rho_{I0}$ with $I=L,C,R$ of the density matrix are
decoupled from the rest and are, therefore, discarded.}
$10N^2\times 10N^2$ which makes direct calculations prohibitive
due to memory and computation time requirements for any realistic
$N$ of the order of 30--40. These problems with the numerical
implementation of the superoperator techniques can be circumvented
by employing iterative methods in which only the procedure/routine
yielding $\mathcal{L}\hat{A}$ for a given $\hat{A}$ is
needed.\cite{pol-jcp-94} Obviously, this does not require the
storage of the whole supermatrix $\mathcal{L}$. On the other hand,
as with any iterative method, the convergence of the iteration
becomes an issue. In Appendix \ref{arnoldi} we give a brief review
of the usage of the Arnoldi iteration in our calculations. Its
intent is to guide the reader through the algorithm so that it can
be reproduced with the help of the mathematical
references.\cite{golub, eirola}

\section{Results}

We now turn to the numerical results for the mean current $I$,
zero-frequency noise $S(0)=S_{ab}(0)$ (for any $a,b$
--- see above), and the Fano factor $F=\tfrac{S(0)}{eI}$ as
functions of the device bias $\eps_b$ for different sets of the
other parameters. First we present a generic plot in the parameter
regime considered by Armour and MacKinnon and comment on the
general features which we can observe in it. We then give a
tentative interpretation of those features supported by
phenomenological arguments and results found in different limiting
cases studied further on. In particular, we consider two specific
limiting cases where at least a partial comparison with
approximate analytic theories can be made, namely (i) the limit of
{\em small damping} which is relevant for the issue of shuttling
and strong inelastic cotunneling and (ii) the limit of {\em weak
inter-dot coupling} which implies in a certain device bias range
the sequential tunneling regime.

\subsection{Generic case}

\begin{figure*}
  \centering
  \includegraphics[width=140mm]{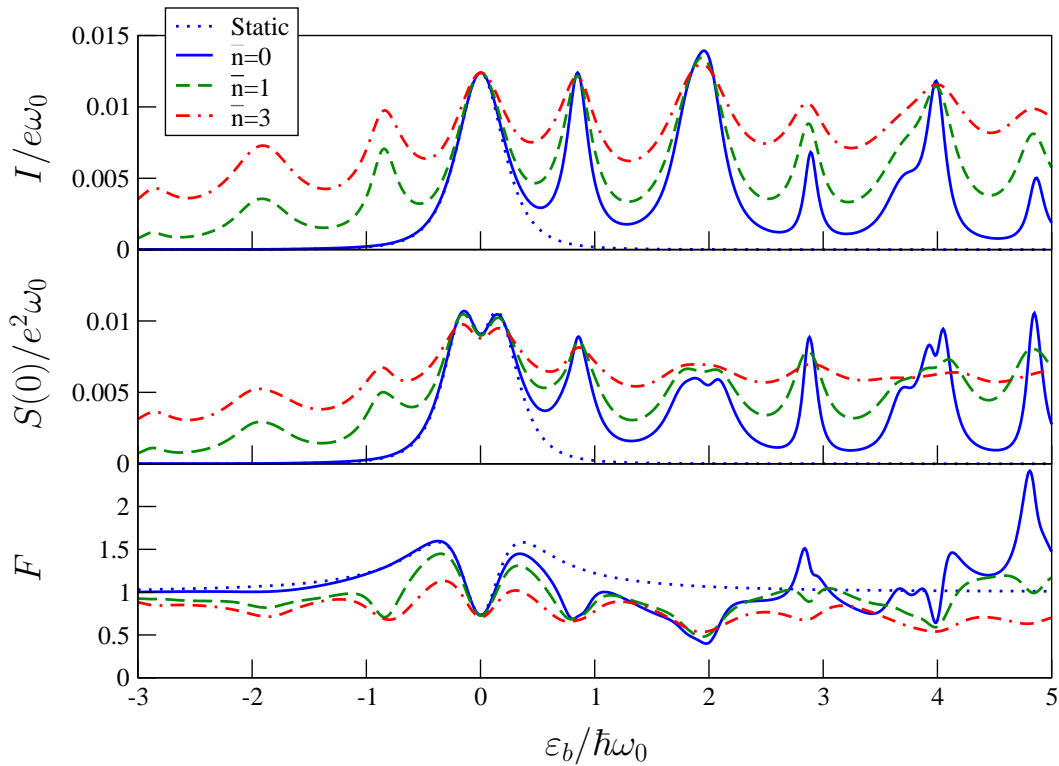}
\caption{The mean current $I$, zero-frequency noise $S(0)$, and
the Fano factor $F$ as functions of the device bias $\eps_b$ for
the static three dot array (dotted line) and the vibrating array
at different temperatures given by the mean oscillator occupation
number $\bar{n}$. The other parameters are
$V_0=0.5\hbar\om_0,\,\alpha=0.2\sqrt{2}\sqrt{m\om_0/\hbar},\,
x_0=(5/\sqrt{2})\sqrt{\hbar/m\om_0},\, \gamma =0.025\om_0,\,
\Gamma=0.05\om_0$ which corresponds to the case studied in
Ref.~\onlinecite{arm-prb-02}, Fig.~6.}\label{generic}
\end{figure*}

In Fig.\ \ref{generic} we plot the mean current, zero-frequency
noise and the Fano factor as functions of the device bias and
temperature for one of the parameter sets considered in
Ref.~\onlinecite{arm-prb-02}. We include non-zero temperature and
extend the device bias range considered
previously\cite{arm-prb-02} to negative values which is relevant
for non-zero temperature.

The dotted lines show the results for the static array. By
applying the theory of Sec.~\ref{theory} to the static array we
found analytic expressions for both the mean current and the Fano
factor which we, however, do not present explicitly here since the
formulas are quite involved. The mean current has a resonant
peak\cite{weg-prb-99} around $\eps_b=0$ while there is a dip in
the noise around $\eps_b=0$ which was also found analytically for
a two dot array by Elattari and Gurvitz.\cite{ela-pla-02} They
attributed the dip to the strong Coulomb interaction on the array.
Our Fano factor shows a crossover from the sub-Poissonian ($F<1$)
dip around $\eps_b=0$ to super-Poissonian ($F>1$) ``shoulders"
starting around $\eps_b\approx\pm V_0e^{-\alpha x_0}$ which
approach the Poissonian limit $F=1$ for large device bias. The
Poissonian limit of the Fano factor for large $\eps_b$ is
understood when one notices that the current in that limit is very
small. Therefore, electrons tunnel through the array sparsely and,
consequently, there is no correlation between successive tunneling
events which form a classical Poisson process with the
(Poissonian) value of the Fano factor $F=1$. While the dip around
zero and the Poissonian limit for large device bias were observed
in the two dot case as well\cite{ela-pla-02} the Fano factor
exceeding one was not present there. We attribute the
super-Poissonian behavior to the (elastic) cotunneling through the
central dot.

Now, let us discuss the results for movable arrays. The
characteristic features are the peaks in current and noise at the
device bias around a non-zero integer multiple of the oscillator
frequency due to electromechanical resonances. The current peaks
at zero temperature (therefore, only for positive multiples of the
frequency) were already observed in previous
works.\cite{arm-prb-02,bra-prb-03} Some of the noise peaks have
further fine structure which is even amplified in the Fano factor
exhibiting a rather complex behavior around the peaks, especially
at low temperature, and showing also strong temperature
dependence.

The zero device bias behavior is clearly governed by the static
array physics which is due to partial decoupling of the electronic
and oscillator degrees of freedom at $\eps_b=0$ when the
electrostatic interaction on the central site
$-\tfrac{\eps_b}{2x_0}\hat{x}\proj{C}{C}$ is turned off. The
remaining interaction stemming from the $x$-dependence of the
hopping amplitudes $t_L(\hat{x}),\,t_R(\hat{x})$ is too weak to
modify the static result in the vicinity of $\eps_b=0$ even for
high temperatures. Some discrepancy between the static and high
temperature dynamic cases around $\eps_b=0$ is found for higher
values of $\alpha\approx 1$ (strictly quantum case from the
oscillator point of view which was previously studied in the
one-dot shuttling setup\cite{nov-prl-03,nov-preprint-04}), yet the
effect is not very pronounced anyway (not shown).

The peaks at non-zero multiples of the oscillator frequency were
already previously attributed to electromechanical
resonances.\cite{arm-prb-02,bra-prb-03} Yet, this explanation is
rather broad and covers a range of processes which can be
responsible for the electronic transport such as cotunneling,
phonon-assisted tunneling, or shuttling occurring around different
resonance peaks.\cite{arm-prb-02, modena} The discrimination
between the different processes is quite complicated since it
cannot be inferred directly from a single $I$ vs.~$\eps_b$ curve.
Either one has to study the dependence of the curves on different
parameters\cite{arm-prb-02} or some other kind of information
about the system must be obtained. A powerful choice is to
calculate and analyze the Wigner distribution functions of the
oscillator in the phase space (possibly
charge-resolved).\cite{bra-prb-03,nov-prl-03, modena} These
characterize the state of the system very well and we will use
them in this study too. However, even though they are an excellent
theoretical tool to study NEMS their connection to data
extractable from a real NEMS experiment is at best remote.
Therefore, diagnostics based on the measurement of the current
statistics is clearly preferable and, therefore, our aim is to
correlate particular features observed in the noise with specific
transport mechanisms within the array as identified by the
theoretical analysis involving also phase space plots.

To achieve this goal we will study different limiting cases in
which particular features of the noise (more precisely of the Fano
factor) are pronounced so that they can be attributed to specific
transport mechanisms. Yet, the results do not allow to associate a
given value of the Fano factor to a specific mechanism. It is more
reading of the whole $I$ vs.~$\eps_b$ curve at least locally
around a peak which gives us the notion of what mechanism(s) are
involved in the transport at that given peak.

As a rule of thumb we can say that the super-Poissonian peaks of
the Fano factor correspond to cotunneling through the central dot.
This statement is supported by the limiting studies discussed
below, and also by the following evidence from Fig.~\ref{generic}.
The peaks only occur for small temperature and disappear with its
increase pointing out to a coherent effect. They also appear
predominantly at odd multiples of the oscillator frequency which
is consistent with the cotunneling picture between the outer dots
excluding the central one due to the energy mismatch. On the other
hand, the dips in the Fano factor curves are due to some form of
the sequential tunneling via the central dot. The most important
aspect is that the process proceeds via a real intermediate state
on the central dot in contrast to the virtual nature of the
cotunneling process. The real sequential process is subject to the
charge conservation which is a strict law strongly suppressing the
Fano factor\cite{suk-prb-01} and causing the dip. The sequential
tunneling picture still involves different mechanisms
distinguished by the detailed state of the oscillator. The
oscillator might be in a general non-equilibrium state during the
tunneling events (this scenario encompasses both the
shuttling\cite{modena} and a general non-equilibrium
oscillator-assisted tunneling\cite{mit-preprint-03} mechanisms) or
it could equilibrate between consecutive tunneling events. The
latter case is studied in detail in subsection~\ref{seq_tun}.

The two charge transfer mechanisms (cotunneling and sequential
tunneling) may coexist, i.e.\ part of the current is carried by
the cotunneling mechanism and the other part by the sequential
tunneling, and their relative weights depend strongly on the
parameters. For example, the transport around $\eps_b\approx
2\hbar\om_0$ is typically governed by shuttling  which results in
the dip while cotunneling is dominant around $\eps_b\approx
3\hbar\om_0$ giving a peak. However, the dip around $\eps_b\approx
\hbar\om_0$ in Fig.~\ref{generic} changes into a clear peak when
$\alpha$ is enlarged up to $\alpha\approx 0.4$ (not shown). This
behavior is still not well understood. Even more complicated is
the behavior around $\eps_b\approx 4\hbar\om_0$ where there is a
dip in the peak. As we show in the next subsection this
corresponds to a fast crossover between the cotunneling and
shuttling transport mechanisms in the vicinity of $\eps_b=4$. In
order to support the above statements for the generic parameters
we study particular limiting cases which enable us to associate
specific features of the Fano factor curves to specific
mechanisms.

\subsection{Small damping: shuttling and strong inelastic
cotunneling}

\begin{figure}
  \centering
  \includegraphics[width=85mm]{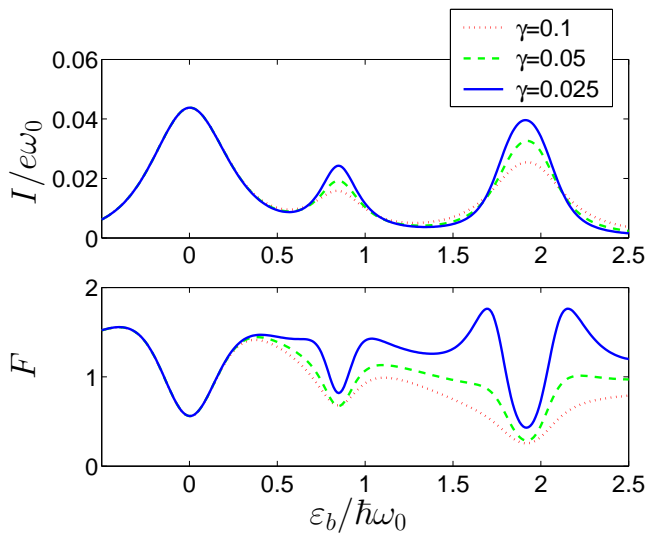}
\caption{The mean current and Fano factor for $V_0=0.76\hbar\om_0,
\alpha=0.28\sqrt{m\om_0/\hbar}, x_0=5\sqrt{\hbar/m\om_0},
\Gamma=0.2\om_0, T=0$ and different values of the damping
coefficient (in units of $\om_0$) corresponding to shuttling
around $\eps_b\approx
\hbar\om_0,2\hbar\om_0$.}\label{shuttlenoise}
\end{figure}

In this section results for small damping case, i.e.\
$\gamma\lesssim\tfrac{I}{e}$ with $I$ a representative value of
the current (given, e.g.\ by its value at the zero device bias
peak), are presented. \footnote{Partial and preliminary results
have already been reported in the shuttling context.\cite{nara,
modena}} First, we focus on the device bias range $\eps_b\approx
0-2.5\hbar\om_0$ where electromechanical instabilities which can
be related to shuttling were inferred indirectly from the behavior
of the mean current,\cite{arm-prb-02} predicted by quasiclassical
studies,\cite{nara} and subsequently directly observed in the
phase space.\cite{modena} The intuition and simple theoretical
estimates (the zero-frequency noise is given by the ratio of the
variance and the square mean of the waiting time between
consecutive loading events of the classical shuttle, see
Eq.~(4.48) in Ref.~\onlinecite{dav-prb-92}) suggest that shuttling
is a low noise phenomenon with the Fano factor close to zero in
the nearly perfectly developed shuttling regime. This was recently
confirmed by more sophisticated calculations for the classical
driven\cite{pis-preprint-04} and quantum\cite{nov-preprint-04}
shuttle in the one-dot setup. In the present, more complicated
setup the shuttling is obscured by competing mechanisms (coherence
between dots, strong Coulomb blockade affecting the whole array)
and we will study the consequence of this fact on the behavior of
the Fano factor.

In Fig.~\ref{shuttlenoise} we show the results for the mean
current and the Fano factor for zero temperature and three
different (small) values of the damping. In
Ref.~\onlinecite{modena} we presented the phase space plots of the
oscillator which we introduce here in more detail later on (see
Eq.~\eqref{wigner} and Fig.~\ref{coexistwigner}). They described a
similar parameter range and showed gradually developing shuttling
around $\eps_b\approx \hbar\om_0,\, 2\hbar\om_0$ with increasing
injection rate $\Gamma$. At these resonance points the current has
peaks moderately changing with the increase of the damping and the
Fano factor has local minima with possible shoulder-like structure
further from the resonance points in case of the smallest damping.
As established more explicitly below, the shoulders are a
signature of coherent processes through the whole array
(cotunneling) and, therefore, are destroyed by the increased
damping.

At the same time the absolute values of the local minima of the
Fano factor at the resonances become deeper by the increased
damping. We conjecture that this somewhat surprising behavior can
also be attributed to the destruction of the quantum coherence and
to the crossover into the non-equilibrium sequential tunneling
regime partially encompassing shuttling. The minimum of the Fano
factor curve starts to increase again with further increase of
damping (not shown) as expected from the classical shuttling
theory. The minimal value of the Fano factor achieved for the
given set of parameters was $F_{\rm min}\approx 0.25$ which
corresponds to a partially developed shuttling regime and was also
confirmed by the phase space pictures (not shown).
\begin{figure}
  \centering
  \includegraphics[width=85mm]{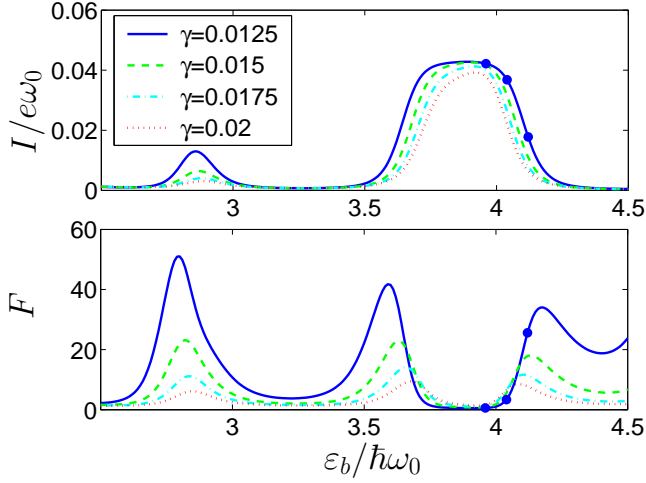}
  \caption{The mean current and Fano factor for $V_0=0.76\hbar\om_0, \alpha=0.28\sqrt{m\om_0/\hbar},
   x_0=5\sqrt{\hbar/m\om_0}, \Gamma=0.2\om_0, T=0$ and different values of the
   damping coefficient (in units of $\om_0$) in the strong inelastic
   cotunneling/shuttling regime. The dots on the curves corresponding
   to $\gamma=0.0125$ denote the points for which the Wigner functions
   in Fig.~\ref{coexistwigner} are plotted.}\label{cotunelnoise}
\end{figure}
\begin{figure}
  \centering
  \includegraphics[width=85mm]{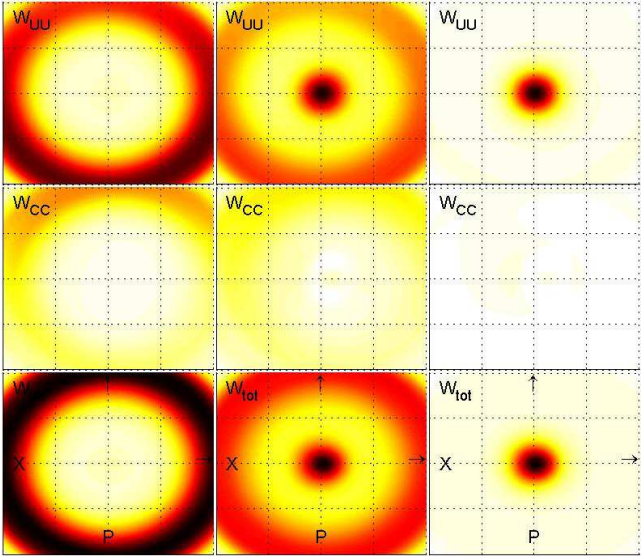}
  \caption{Phase space representation of the oscillator around the
    transition from the shuttling to the strong inelastic cotunneling
    regime at $\eps_b/\hbar\om_0=3.96,4.04,4.12$, respectively
    (columns from the left to the right). The respective rows show
    the Wigner distribution functions  for the empty ($W_{UU}$) or occupied
    ($W_{CC}$) central dot, and the sum of the two ($W_{\rm tot}=W_{UU}+W_{CC}$)
    in the oscillator phase space (horizontal axis -- coordinate in units of
    $\sqrt{\hbar/m\om_0}$,  vertical axis -- momentum in $\sqrt{\hbar m\om_0}$,
    the grid is at 2.5 in the dimensionless units).
    The other parameters are: $V_0=0.76\hbar\om_0,\alpha=0.28\sqrt{m\om_0/\hbar},
    x_0=5\sqrt{\hbar/m\om_0}, \gamma=0.0125\om_0, \Gamma=0.2\om_0,
    T=0$. The parameters correspond to the dots in Fig.~\ref{cotunelnoise}.
    The Wigner functions are normalized within each column.}\label{coexistwigner}
\end{figure}

Next, we focus on the range $\eps_b \approx
2.5\hbar\om_0-4.5\hbar\om_0$ involving two current peaks around
$\eps_b \approx 3\hbar\om_0,\, 4\hbar\om_0$. As we already
mentioned in the generic case the peak around $\eps_b \approx
3\hbar\om_0$ corresponds to cotunneling while the behavior around
$\eps_b \approx 4\hbar\om_0$ is given by a complicated interplay
between both mechanisms (cotunneling and sequential tunneling).
With lower damping the differences in the Fano factors of the two
mechanisms become more pronounced as we show in
Figs.~\ref{cotunelnoise} and \ref{coexistwigner}. In
Fig.~\ref{cotunelnoise} the mean current and the Fano factor as
functions of the device bias $\eps_b$ are depicted for several
(small) values of the damping. We see the strong damping
dependence of the mean current and the Fano factor around
$\eps_b\approx 3\hbar\om_0$ and in the ``shoulder region" around
$\eps_b\approx 4\hbar\om_0$. On the other hand the mean current as
well as the Fano factor do not depend strongly on the damping in
the close vicinity of $\eps_b\approx 4\hbar\om_0$.

We attribute the first type of behavior to cotunneling. It is
manifested by a strong damping dependence of the current and the
Fano factor, the Fano factor reaches very high values of the order
of $F\approx 50$ for small enough damping. The threshold for the
quasi-divergent behavior of the Fano factor is roughly
$\gamma_{\rm thresh}\approx \tfrac{I}{e}$; for the damping below
this threshold the Fano factor starts to increase. We want to
point out that a giant (divergent) super-Poissonian noise was
theoretically predicted for a quantum dot system in the (strong
inelastic) cotunneling regime analogous to ours by Sukhorukov et
al.\cite{suk-prb-01} The divergence of the current noise is
explained as a slow switching between two or more current channels
carrying different currents. We expect that the different current
channels are formed from different resonant quantum states
connecting the left and right dots in the cotunneling regime. Due
to the small damping rate the switching between those channels is
slow giving rise to the highly super-Poissonian noise.

We also observed a quasi-divergent Fano factor (up to $F\approx
600$) around the shuttling instability transition point in the
quasiclassical limit of the original one-dot shuttle
setup.\cite{nov-preprint-04} The explanation of the divergence is
again the same, i.e.\ the slow switching between different current
channels. Contrary to the present case the two channels of the
one-dot setup are both given by real sequential tunneling
processes via the dot differing just by the state of the
oscillator (equilibrated vs.~shuttling). The switching rate
between the channels can be calculated semi-analytically thus
quantitatively confirming the proposed
mechanism.\cite{don-preprint-04} In the three-dot case the
semi-analytic theory would be much more complicated and we do not
attempt it. A similar mechanism for the quasi-divergent Fano
factor in a single-electron-transistor NEMS was also proposed
recently by Blanter et al.\cite{bla-preprint-04}

Further insight to the details of the microscopic transport
mechanism can be gained by studying the Wigner functions which
describe the oscillator phase space quasiprobability
distributions. We define Wigner functions of the unoccupied
($W_{UU}$), occupied ($W_{CC}$) central dot and their sum ($W_{\rm
tot}$), respectively:
\begin{equation}\label{wigner}
\begin{split}
    W_{UU}(X,P) &= \int_{-\infty}^{\infty}\frac{dy}{2\pi}\,e^{iPy}\\
    &\times\bigl\langle X-\frac{y}{2}\,\big|\big(\hat{\rho}_{00}^{\rm stat}+\hat{\rho}_{LL}^{\rm stat}
    +\hat{\rho}_{RR}^{\rm stat}\big)\big|X+\frac{y}{2}\bigr\rangle\ ,\\
    W_{CC}(X,P) &= \int_{-\infty}^{\infty}\frac{dy}{2\pi}\,e^{iPy}\bigl\langle
    X-\frac{y}{2}\,\big|\hat{\rho}_{CC}^{\rm
    stat}\big|X+\frac{y}{2}\bigr\rangle\ ,   \\
    W_{\rm tot}(X,P)&=W_{CC}(X,P)+W_{UU}(X,P)\ .
\end{split}
\end{equation}
The behavior in the close vicinity of $\eps_b \approx 4\hbar\om_0$
characterized by a weak damping dependence of the mean current and
the Fano factor (of the order of $1$) seen in
Fig.~\ref{cotunelnoise} is characteristic of shuttling. It is
confirmed directly by the phase space plots in
Fig.~\ref{coexistwigner} where the crossover from the
predominantly shuttling transport at $\eps_b=3.96\hbar\om_0$ to
the cotunneling regime at $\eps_b=4.12\hbar\om_0$ is shown.  The
shuttling is evidenced by the asymmetric Wigner distributions of
the occupied or empty central dot $W_{CC}, W_{UU}$, respectively
(first column). The cotunneling manifests itself by the striking
absence of any occupation of the central dot (last column) which
proves the virtual nature of the transport in that case.

\subsection{Weak inter-dot coupling: sequential tunneling assisted by equilibrated oscillator}
\label{seq_tun}

Here we examine the behavior of the system in the weak tunneling
regime, i.e.\ when the hopping elements
$t_L(\hat{x}),\,t_R(\hat{x})$ coupling the adjacent dots in the
array are small and the time scale between tunneling events is
correspondingly the largest in the problem. In this limit the
phonon subsystem gets equilibrated between the consecutive
tunneling events and the distribution of the oscillator and bath
may be taken at equilibrium corresponding to the appropriate
electronic state. We can then solve the GME \eqref{GME1} using
perturbation theory keeping only the lowest order terms in the
bare hopping parameter $V_0$ which turns out to be equivalent to
the $P(E)$-theory.\cite{nazarov} The coherence of the electron
transfer process from the left to the right dot is broken during
the transfer by the long enough interaction with the phonon
subsystem acting as equilibrated thermal bath and, therefore, the
resulting picture is just sequential tunneling (ST) via the
central dot, at least in  the device bias range where the above
assumptions hold. We defer a more detailed discussion until the
end of this subsection where the assumptions will be reexamined
and their validity clarified in view of the obtained results.

\begin{figure}[htbp]
\begin{center}
\includegraphics[width=0.4\textwidth]{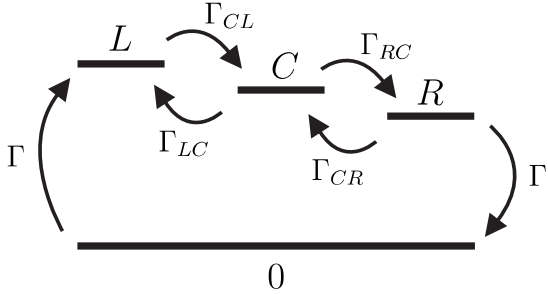}
\caption{The four states $0$ (device empty), $L$ (left dot
occupied), $C$ (center dot occupied), $R$ (right dot occupied) and
the transition rates as described by the Markov process given by
the transition matrix (\ref{stomatrix}).} \label{markovchain}
\end{center}
\end{figure}

When we carry out the approximate solution of Eq.~\eqref{GME1} in
the lowest order in $V_0$ as described in Appendix
\ref{P(E)-theory} we obtain the rate equation \eqref{rate_eq}
describing a classical Markov process of the sequential electron
transfer between the 4 states which is depicted in
Fig.~\ref{markovchain}. After introducing the vector of occupation
probabilities $\mathbf{p}=[P_0, P_L, P_C, P_R]^T$ the equation can
be rewritten in the matrix form $\dot{\mathbf{p}}=\mathbf{M p}$
with the transition matrix
\begin{equation}\label{stomatrix}
 \mathbf{M}=\begin{pmatrix}
  -\Gamma & 0 & 0 & \Gamma\\
  \Gamma & -\Gamma_{CL} & \Gamma_{LC} & 0 \\
  0 & \Gamma_{CL} & -(\Gamma_{LC}+\Gamma_{RC}) &\Gamma_{CR}  \\
  0 & 0 & \Gamma_{RC}& -(\Gamma_{CR}+\Gamma)
\end{pmatrix}.
\end{equation}
The rates entering the matrix are calculated as functions of the
model parameters from the microscopic $P(E)$-theory and the
results are given in Appendix \ref{P(E)-theory},
Eqs.~\eqref{rates_F_function},~\eqref{FG_functions},~\eqref{F_function}.
The stationary state $\mathbf{p}^{\mathrm{stat}}$ satisfying
$\mathbf{M p}^{\mathrm{stat}}=\mathbf{0}$ is found to be
\begin{equation}
\mathbf{p}^{\mathrm{stat}} = N \begin{pmatrix}
  \Gamma_{CL}\Gamma_{RC}\\
  \Gamma_{RC}\Gamma+\Gamma_{LC}(\Gamma_{CR}+\Gamma)\\
  \Gamma_{CL}(\Gamma_{CR}+\Gamma)\\
  \Gamma_{CL}\Gamma_{RC}
\end{pmatrix}
\end{equation}
with the normalization constant
$N=\big(\Gamma_{RC}\Gamma+\Gamma_{LC}(\Gamma_{CR}+\Gamma)+\Gamma_{CL}
(\Gamma_{CR}+2\Gamma_{RC}+\Gamma)\big)^{-1}$.

To calculate the mean current and, in particular, the current
noise one can proceed following two possible equivalent ways which
parallel in close analogy the two methods used in subsections
\ref{qrth} and \ref{mcd}. In the first method found in
Refs.~\onlinecite{dav-prb-92}, \onlinecite{her-prb-93},
\onlinecite{kie-prb-03} one defines an effective operator for the
current running between, e.g.\ $L$ and $C$ by
\begin{equation}\label{current_seqtun}
\mathbf{I}_{CL} = e\begin{pmatrix}
  0 & 0 & 0 & 0\\
  0 & 0 & -\Gamma_{LC} & 0\\
  0 & \Gamma_{CL} & 0 & 0\\
  0 & 0 & 0 & 0
\end{pmatrix},
\end{equation}
and together with the definition of the trace of a vector
$\mathbf{v}$ as the sum of its elements, i.e.
$\mathrm{Tr}\,\mathbf{v} = \sum_j v_j$, the mean steady state
current $I$ reads
\begin{equation}
I=\langle \mathbf{I}_{CL} \rangle =
\mathrm{Tr}(\mathbf{I}_{CL}\mathbf{p}^{\mathrm{stat}}) =
N\Gamma\Gamma_{CL}\Gamma_{RC}. \label{current4}
\end{equation}

Using the current operator we consider the current-current
correlation function
\begin{equation}
C_{CL,CL}(\tau) =\langle
\mathbf{I}_{CL}(\tau)\mathbf{I}_{CL}(0)\rangle-\langle
\mathbf{I}_{CL}\rangle^2,
\end{equation}
with the current-current correlator given by Hershfield et
al.\cite{her-prb-93} as
\begin{equation}
\begin{split}
\langle \mathbf{I}_{CL}(\tau)\mathbf{I}_{CL}(0)\rangle
&=\theta(\tau)\,\mathrm{Tr}\bigl(\mathbf{I}_{CL}\mathbf{T}(\tau)\mathbf{I}_{CL}
\mathbf{p}^{\mathrm{stat}}\bigr)\\
&+\theta(-\tau)\,\mathrm{Tr}\bigl(\mathbf{I}_{CL}\mathbf{T}(-\tau)\mathbf{I}_{CL}
\mathbf{p}^{\mathrm{stat}}\bigr)\\
&+e\delta(\tau)\mathrm{Tr}\bigl|\mathbf{I}_{CL}
\mathbf{p}^{\mathrm{stat}}\big|
\end{split}
\label{correlator}
\end{equation}
with the time propagator $\mathbf{T}(\tau)=\exp(\mathbf{M}\tau)$
and $\mathrm{Tr}|\mathbf{v}|=\sum_j|v_j|$. This fully classical
formula bears some formal resemblance to the quantum case
\eqref{qrt} but there is an important difference in the presence
of the $\delta$-function term in \eqref{correlator}. While the
first two terms of \eqref{correlator} correspond to correlations
between different tunneling events, the third term describes the
self-correlation of a single tunneling event within the classical
description.  The self-correlation term cannot be derived within
the rate equation formalism and was inserted by hand into the
noise formula of Ref.~\onlinecite{her-prb-93} based on the results
of the previous more microscopic study.\cite{dav-prb-92} Following
the same line of arguments as in Sec.~\ref{qrth} we get the
following expression for the Fano factor $F=\tfrac{S(0)}{eI}$
\begin{equation}\label{fano}
F=\frac{-2\mathrm{Tr}(\mathbf{I}_{CL}\mathbf{Q}\mathbf{M}^{-1}
\mathbf{Q}\,\mathbf{I}_{CL}\mathbf{p}^{\mathrm{stat}})+e\mathrm{Tr}\bigl|\mathbf{I}_{CL}
\mathbf{p}^{\mathrm{stat}}\big|} {e\langle\mathbf{I}_{CL}\rangle}
\end{equation}
with the projector $\mathbf{Q}=\mathbf{1}-\mathbf{p}^{\rm
stat}\otimes[1,1,1,1],\ \mathbf{Q}^2=\mathbf{Q}$. Therefore, the
Fano factor is determined by the pseudoinverse of the transition
matrix $\mathbf{Q}\mathbf{M}^{-1}\mathbf{Q}$ in analogy with the
quantum-mechanical case.

Exactly the same formula for the Fano factor can be obtained by
employing the full counting statistics approach analogous to the
calculations in Sec.~\ref{mcd} applied to the classical rate
equation. To calculate the noise one has to introduce the counting
variable $n$ describing the number of electrons that tunneled
across a chosen junction, e.g.\ the LC-junction between the left
and the central dot. Since in the present setup electrons can
tunnel in the backwards direction, i.e.\ from the central dot to
the left dot (see Fig.~\ref{markovchain}), $n$ can become negative
as well. This technical detail slightly modifies the derivation
which, however, closely follows the previous lines. We start with
Eq.~\eqref{macdonald} where the probability that $n$ electrons
tunneled across the LC junction (positive $n$ corresponds to the
left-to-center direction)
$P_n(t)=P_0^{(n)}(t)+P_L^{(n)}(t)+P_C^{(n)}(t)+P_R^{(n)}(t)$ is
determined by the $n$-resolved form of the rate equation
\begin{equation}
\begin{split}
\dot{P}_0^{(n)} &= -\Gamma P_0^{(n)} + \Gamma P_R^{(n)}\\
\dot{P}_L^{(n)} &= \Gamma P_0^{(n)} - \Gamma_{CL} P_L^{(n)} + \Gamma_{LC} P_C^{(n+1)} \\
\dot{P}_C^{(n)} &= \Gamma_{CL}
P_L^{(n-1)}-(\Gamma_{LC}+\Gamma_{RC})
P_C^{(n)}+\Gamma_{CR} P_R^{(n)}\\
\dot{P}_R^{(n)} &=\Gamma_{RC}
P_C^{(n)}-(\Gamma_{CR}+\Gamma)P_R^{(n)}
\end{split}
\end{equation}
which is an intuitive generalization of the original rate equation
\eqref{rate_eq} obtained by including the transferred charge
statistics across the LC junction, see Fig.~\ref{markovchain}.
Performing the calculation of the noise from \eqref{macdonald} in
the spirit of Sec.~\ref{mcd} we come to the formula \eqref{fano}
again. We want to stress that using this second way of derivation
gives us the entire formula with the self-correlation term and
even the definition of the current operator \eqref{current_seqtun}
appearing naturally in the course of the derivation. In this sense
the intuitive generalization of the rate equation incorporating
the transferred charge resolution yields the full microscopic
description of the whole process (contrary to the bare rate
equation) and no heuristic arguments are necessary to get the
self-correlation term.

\begin{figure}
  \centering
  \includegraphics[width=85mm]{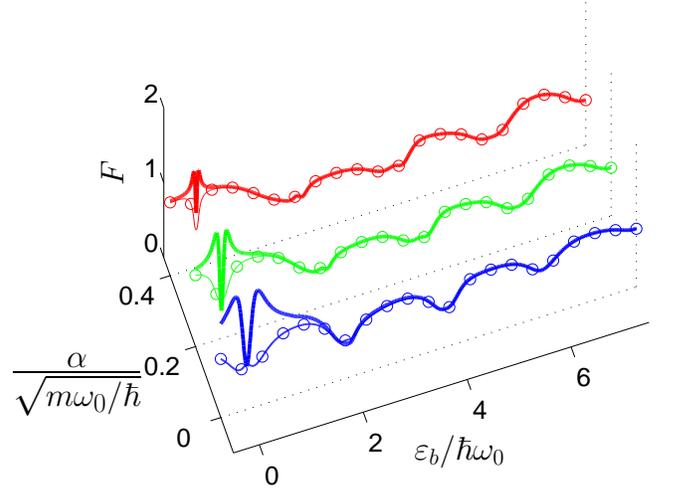}
\caption{The Fano factor in the two-state sequential tunneling
limit (zero temperature, large $\Gamma$). The thick line is the
computed Fano factor while the thin lines with circles are given
by the formula $n_C^2+(1-n_C)^2$ where $n_C$ are the occupation of
the central dot. The collapse of the two curves marks the
sequential tunneling region. The values of the other parameters
are $V_0=0.1\hbar\om_0,\, x_0=5\sqrt{\hbar/m\om_0},\, \gamma
=0.1\om_0,\,\Gamma=0.1\om_0,\, T=0$.}\label{ST_check}
\end{figure}

For the process determined by the rate matrix \eqref{stomatrix}
the Fano factor can be rather easily evaluated
analytically.\cite{christian} The resulting expression is,
however, complicated and will not be given here. In the limit when
$\Gamma \gg \Gamma_{CL},\Gamma_{LC},\Gamma_{RC},\Gamma_{CR}$ only
the left or the central dot are occupied since the right dot and
unoccupied state are immediately emptied in favor of the left dot.
Due to the zero occupation of the right dot, the rate
$\Gamma_{CR}$ despite its non-zero value drops out from the
expressions for the stationary probability distribution, mean
current, and Fano factor. If, moreover, the temperature is zero we
expect the rate $\Gamma_{LC}$ to vanish (for $T=0$ only the
positive device bias range $\eps_b>0$ is interesting from the ST
point of view) and the stationary probability, mean current and
Fano factor assume the well-known form for a two-state
process\cite{dav-prb-92,bla-phr-00}
\begin{subequations}\label{seq_tun_Fano}
\begin{align}
\mathbf{p}^{\mathrm{stat}}_{\,\Gamma\to\infty,T=0}&=\frac{1}{\Gamma_{CL}+\Gamma_{RC}}
\begin{pmatrix}0 \\\Gamma_{RC}\\\Gamma_{CL}\\0\end{pmatrix}\\
I_{\,\Gamma\to\infty,T=0} &=
\frac{\Gamma_{CL}\Gamma_{RC}}{\Gamma_{CL}+\Gamma_{RC}}\\
F_{\,\Gamma\to\infty,T=0}&=\frac{\Gamma_{CL}^2+
\Gamma_{RC}^2}{(\Gamma_{CL}+\Gamma_{RC})^2}\ .
\end{align}
\end{subequations}
As a consequence of these relations the Fano factor can be
expressed in the limit $\Gamma\to\infty,T=0$ in terms of, e.g.,
the stationary occupation $n_C=P_C^{\rm stat}$ of the central dot
as $F=n_C^2+(1-n_C)^2$. This is an identity relating the Fano
factor and the central dot occupation in the ST regime regardless
of the particular values of the rates provided that the above
assumptions are fulfilled.

In Fig.~\ref{ST_check} we show the Fano factor as a function of
the device bias for small $V_0$, zero temperature, and three
different values of $\alpha$  calculated numerically by the method
described in Sec.~\ref{numerics}. We expect the system to be in
the two-state ST regime described above. The thick lines are the
Fano factor calculated directly while the thin lines with circles
show the quantity $n_C^2+(1-n_C)^2$ with $n_C$ being the
occupation of the central dot calculated from the numerical
evaluation of the full $\hat{\rho}^{\rm stat}$. We see a nice
collapse of the two curves for roughly $\eps_b\gtrsim
1.5\hbar\om_0$ (depending slightly on the value of $\alpha$). The
collapse marks the two-state ST region. The discrepancy around $0<
\eps_b\lesssim 1.5\hbar\om_0$ is due to cotunneling processes
prevailing over the ST ones in that region of $\eps_b$. The
electromechanical coupling terms are proportional to $\eps_b$ and
$V_0$ and, therefore, the heat bath consisting of the mechanical
degrees of freedom gets almost decoupled in the ST regime (small
$V_0$) at small $\eps_b$ and does not suffice to break the
coherence of the cotunneling processes.

\begin{figure}
  \centering
  \includegraphics[width=85mm]{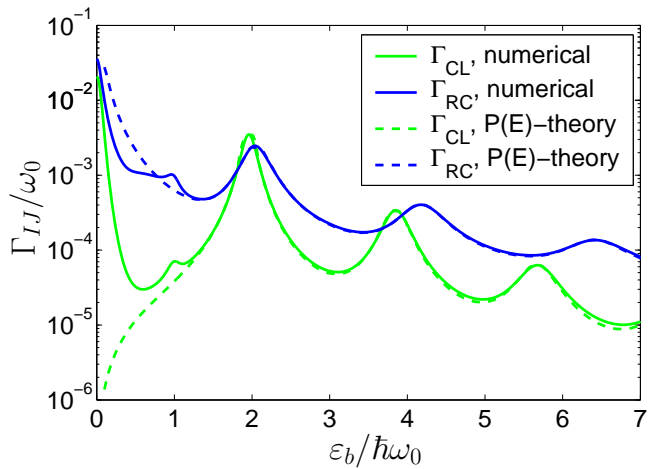}
\caption{Comparison between the numerical rates and the ones
calculated by the $P(E)$-theory for
$V_0=0.1\hbar\om_0,\,\alpha=0.2\sqrt{m\om_0/\hbar}, \,
x_0=5\sqrt{\hbar/m\om_0},\, \gamma =0.1\om_0,\,\Gamma=0.1\om_0,\,
T =0$. The numerical rates are calculated assuming that the
two-state sequential tunneling picture holds which is only true
for $\eps_b\gtrsim 1.5\hbar\om_0$, see Fig.~\ref{ST_check}. In
that region the two results match almost
perfectly.}\label{ST_rates}
\end{figure}

We have thus verified that the identity implied by the two-state
ST process is satisfied by the numerical results. While it helped
us to identify the region of ST, however, the mentioned identity
does not depend on the values of the rates. In the next step we
calculate the values of the rates $\Gamma_{CL},\Gamma_{RC}$ from
the numerical results for the mean current and occupation of the
central dot or Fano factor by inverting Eqs.~\eqref{seq_tun_Fano},
plot them in Fig.~\ref{ST_rates}, and compare with the rates
calculated semi-analytically according to the $P(E)$-theory
presented in Appendix \ref{P(E)-theory}. We see a nearly perfect
match between the two approaches in the regime of the two-state
ST. The numerical rates were calculated using
Eqs.~\eqref{seq_tun_Fano} in the whole range of $\eps_b$ and,
therefore, do not represent the correct rates in the cotunneling
dominated regime $\eps_b\lesssim 1.5\hbar\om_0$. The
semi-analytical rates also confirm the cause of the ST mechanism
breakdown discussed above. The $\Gamma_{CL}$ rate yielding the
bottleneck of the ST current essentially vanishes below the ST
threshold and higher order processes in $V_0$ (cotunneling) take
over.

\begin{figure}
  \centering
  \includegraphics[width=85mm]{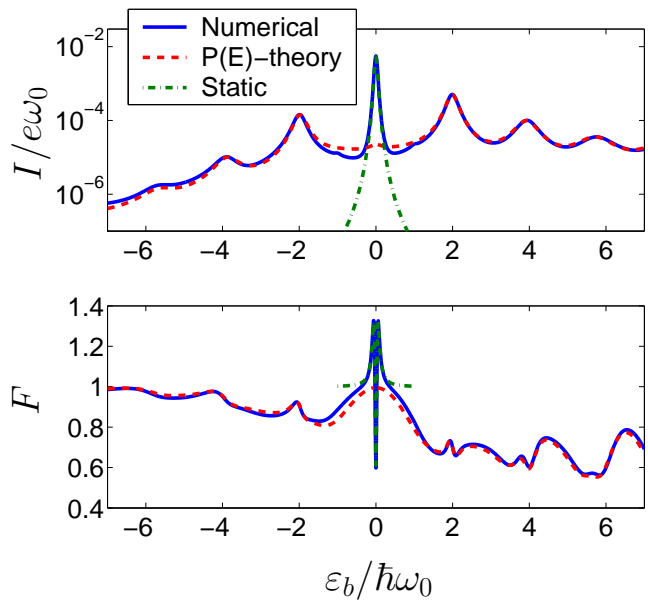}
\caption{Comparison between the numerics and the $P(E)$-theory
based sequential tunneling picture for
$V_0=0.05\hbar\om_0,\,\alpha=0.2\sqrt{m\om_0/\hbar}, \,
x_0=5\sqrt{\hbar/m\om_0},\, \gamma =0.1\om_0,\,\Gamma=0.1\om_0,\,
\bar{n} = 1$. Due to the non-zero temperature the two-state model
considered in Fig.~\ref{ST_check}, \ref{ST_rates} had to be
extended and there is a new sequential tunneling region also for a
negative bias. We observe a nearly perfect match between the two
approaches for $|\eps_b|\gtrsim 1.5\hbar\om_0$. The behavior
around $\eps_b=0$ is clearly governed by the physics of the static
array since the oscillator is largely decoupled from the
electronic degrees of freedom.}\label{ST_results}
\end{figure}

We show a representative plot of the general ST results without
the assumptions
$T=0,\,\Gamma\gg\Gamma_{CL},\Gamma_{LC},\Gamma_{RC},\Gamma_{CR}$
in Fig.~\ref{ST_results}. The comparison between the numerically
calculated and semi-analytical results is shown for both the mean
current (log scale) and the Fano factor. Since the temperature is
non-zero there is a new ST region for a negative bias. We see a
good match between the two approaches in the bias range $|\eps_b|
\gtrsim 1.5\hbar\om_0$. The fine structure around $\eps_b$ being
an even multiple of the oscillator frequency is given by the
interplay between the values of different tunneling rates in those
regions similar to the switching of the relative magnitude of
$\Gamma_{CL}$ and $\Gamma_{RC}$ in Fig.~\ref{ST_rates}. The
behavior around $\eps_b=0$ is clearly given by the physics of the
static array also shown in the figure so that there are only small
regions around $\eps_b=\pm\hbar\om_0$ which are not covered either
by the ST or static picture. To summarize, we have shown that the
electronic transport through the array in the small $V_0$ limit
can be successfully described (in the device bias range $|\eps_b|
\gtrsim 1.5\hbar\om_0$) by the ST theory with the transfer rates
determined semi-analytically by the microscopic $P(E)$-theory.

\section{Conclusions}
We have developed theoretical techniques to evaluate the
zero-frequency current noise in nanoelectromechanical systems. Two
parallel lines have been developed: (i) Quantum regression theorem
(QRT) and (ii) Full counting statistics (FCS).  QRT has the
advantage of being applicable to {\it any} correlation functions
involving exclusively system operators, while FCS gives perhaps a
more direct access to the current noise, but, on the other hand,
other correlation functions cannot directly be accessed with it.
We have demonstrated the equivalence of the two approaches for the
model considered in this work, but we emphasize that the
equivalence is critically dependent on whether charge conserving
approximations are used.  The three-dot model considered in this
paper has a rich phenomenology allowing one to study the effect of
the internal coherence of the electronic states, and by tuning the
system parameters we can study the transition from a co-tunneling
dominated regime to a sequential tunneling regime.  The
generalized master equations studied in this paper involve large
matrices, and we have discussed in detail the numerical schemes
that are needed in their solution.  In certain limiting cases
approximate (semi)analytic theories can be developed, and we have
found an excellent agreement with the full numerics.  We have
interpreted the computed current and noise curves in terms of
physical concepts, and gained an understanding of when one can
expect either sub- or super-Poissonian behavior. We believe that a
successful interpretation of numerical results requires a
simultaneous analysis of several quantities such as mean current,
Fano factor and Wigner distributions.

There are several lines along which the present approach can be
continued.  An interesting and important issue concerns the
finite-frequency noise, and we are presently examining extensions
of our theory in that direction.  Spin-degree of freedom has been
neglected in our calculations, and more work in that direction is
called for.  We have pointed out certain restrictions in the
derivation of the generalized master equations, and one should
look carefully at effects of (i) a more realistic confining
potential, (ii) the interplay of the two different baths, and
(iii) issues related to charge conservation.  We also expect to
get inspiration from experimental studies of quantum shuttles,
which we hope are soon realized.\cite{ble-phr-04}

\begin{acknowledgments}
We thank A.~Armour, T.~Brandes, T.~Eirola, K.~Flensberg, G.
Kie{\ss}lich, A.~Pomyalov, and A.~Wacker for stimulating
discussions and comments. We are very grateful to A.~Donarini for
sharing numerical codes with us and illuminating discussions. This
work was supported by the Oticon Foundation (C.F.) and the grant
202/01/D099 of the Czech grant agency (T.N.) which we gratefully
acknowledge.
\end{acknowledgments}

\appendix

\section{Arnoldi iteration}\label{arnoldi}

The key concept of the Arnoldi iteration is the construction of
the Krylov subspace
$\mathcal{K}_j(\mathcal{L},\mathbf{x}_0)=\mathrm{span}(\mathbf{x}_0,\,
\mathcal{L}\mathbf{x}_0,\,\mathcal{L}^2\mathbf{x}_0,\dots,\mathcal{L}^{j-1}\mathbf{x}_0)$
for a chosen initial supervector $\mathbf{x}_0$ and successively
the computation of an orthonormal basis
$\mathbf{Q}_j=[\mathbf{q}_1,\dots,\mathbf{q}_j]$ in it by the
Gram-Schmidt orthogonalization. In the orthogonalization process
defined by the recurrence relation
\begin{equation}
\begin{split}
    \mathbf{q}_1 &= \frac{\mathbf{x}_0}{\norm{\mathbf{x}_0}}\ , \\
    \mathbf{q}_{k+1} &= \frac{\mathcal{L}\mathbf{q}_k
    - \sum_{i=1}^{k} (\mathbf{q}_i^{\dag}\cdot\mathcal{L}\mathbf{q}_k)\,
    \mathbf{q}_i}{\norm{\mathcal{L}\mathbf{q}_k
    - \sum_{i=1}^{k} (\mathbf{q}_i^{\dag}\cdot\mathcal{L}\mathbf{q}_k)\,
    \mathbf{q}_i}}
    \ ,\ \text{ for  } k = 1,\dots j
\end{split}
\end{equation}
a complex upper $(j+1)\times j$ Hessenberg matrix
\begin{equation}
\mathbf{H}_j=\begin{pmatrix}
  h_{1,1} & h_{1,2} & h_{1,3} & h_{1,4}& \dots & h_{1,j}\\
  h_{2,1} & h_{2,2} & h_{2,3} & h_{2,4}& \dots & h_{2,j}\\
  0       & h_{3,2} & h_{3,3} & h_{3,4}& \dots & h_{3,j}\\
  0       & 0       & h_{4,3} & h_{4,4}& \dots & h_{4,j}\\
  \vdots  & \vdots  & \vdots  & \ddots & \ddots& \vdots \\
  0       & 0       & 0       & 0      & \dots     & h_{j+1,j}
\end{pmatrix} \in\mathbb{C}^{(j+1)\times j}
\end{equation}
is recorded with the elements
$h_{i,k}=\mathbf{q}_i^{\dag}\cdot\mathcal{L}\mathbf{q}_k$, for
$i=1,\dots,k \leq j$ and
$h_{k+1,k}=\norm{\mathcal{L}\mathbf{q}_k-\sum_{i=1}^k
h_{i,k}\,\mathbf{q}_i}$ for $k=1,\dots,j$. It enters the following
important relation
\begin{equation}\label{identity}
    \mathcal{L}\mathbf{Q}_j=\mathbf{Q}_{j+1}\cdot\mathbf{H}_j\ .
\end{equation}
Before proceeding we stress the main feature of the iterative
Krylov subspace methods which consists in the fact that the
dimension of the Krylov space is considerably smaller than the
dimension of the original space in which (truncated) $\mathcal{L}$
acts ($j=20$ in our calculations compared to the dimension of
$10N^2\approx 20000$ of the relevant part of the truncated
superspace). The required operations like finding the null space
or the pseudoinverse of $\mathcal{L}$ are performed approximately
in the Krylov subspace only (in the sense specified below) which
makes them very fast. These fast operations are then iterated in
order to achieve the solution of the original problem.

The first task is to calculate the stationary density matrix
$\hat{\rho}^{\rm stat}$ from Eq.\ \eqref{stat}. This means we are
looking for the unique null vector of the superoperator
$\mathcal{L}$. We choose an arbitrary initial vector
$\mathbf{x}_0$ (whose choice can be motivated by a physical guess
of the stationary state to improve the convergence) and construct
the Krylov subspace $\mathcal{K}_j(\mathcal{L},\mathbf{x}_0)$ for
a fixed small $j$. Then we look for a vector
$\mathbf{x}=\mathbf{Q}_j\cdot\xi,\ \xi=(\xi_1,\dots,\xi_j)^T,\
\norm{\xi}=1$ in the subspace which minimizes the norm
$\norm{\mathcal{L}\mathbf{x}}$ in order to approximate the null
vector. Using \eqref{identity} the problem can be reformulated as
\begin{equation}
\min_{\norm{\xi}=1}\,\norm{\mathcal{L}\mathbf{Q}_j\cdot\xi}=
\min_{\norm{\xi}=1}\,\norm{\mathbf{Q}_{j+1}\cdot\mathbf{H}_j\cdot\xi}=
\min_{\norm{\xi}=1}\,\norm{\mathbf{H}_j\cdot\xi}
\end{equation}
due to the property
$\norm{\mathbf{Q}_{j+1}\cdot\mathbf{u}}=\norm{\mathbf{u}}$ for an
arbitrary vector $\mathbf{u}=(u_1,\dots,u_{j+1})^T$.

The last step leaves us with a problem of minimizing the norm in a
$j$-dimensional space spanned by the columns of $\mathbf{H}_j$
which can be solved by performing the singular value decomposition
$\mathbf{H}_j=\mathbf{U}\mathbf{\Sigma}\mathbf{V}^{\dag}$ of the
rectangular matrix $\mathbf{H}_j$.
$\mathbf{U}\in\mathbb{C}^{(j+1)\times (j+1)}$ and
$\mathbf{V}\in\mathbb{C}^{j\times j}$ are unitary matrices whereas
$\mathbf{\Sigma}=\left(\begin{smallmatrix}\sigma_1 &
  & \\  &\sigma_2 &  \\  &  & \ddots \end{smallmatrix}\right)
\in\mathbb{C}^{(j+1)\times j}$ is diagonal with positive
$\sigma_k$'s being the eigenvalues of
$\sqrt{\mathbf{H}_j^{\dag}\cdot\mathbf{H}_j}$ sorted in the
descending order,\cite{golub} i.e.\ $\sigma_1 \geq \sigma_2
\geq\dots\sigma_j\geq 0$. The norm $\norm{\mathbf{H}_j\cdot\xi}=
\norm{\mathbf{\Sigma}\mathbf{V}^{\dag}\cdot\xi}$ is minimized by
choosing for $\xi$ the last column of $\mathbf{V}$ belonging to
the smallest singular value $\sigma_j$ of $\mathbf{H}_j$, i.e.\
$\xi=\mathbf{v}_j$. The vector
$\mathbf{x}=\mathbf{Q}_j\cdot\mathbf{v}_j$ is then an approximate
null vector of $\mathcal{L}$. If the norm
$\norm{\mathcal{L}\mathbf{x}} > tol$ one replaces the initial
guess $\mathbf{x}_0$ by $\mathbf{x}$ and repeats the procedure.
The tolerance was chosen as $tol=10\,\epsilon \,
\norm{\mathcal{L}}$ with $\epsilon$ being the machine precision
and the norm of the Liouvillean was estimated\footnote{T. Eirola,
private communication} as $\norm{\mathcal{L}}=\exp(N/\log(N))$.

To ensure the convergence of the iteration it may be necessary to
use preconditioning, i.e.\ one solves
$\widetilde{\mathcal{L}}\mathbf{x}=0$ where
$\widetilde{\mathcal{L}}=\mathcal{M}^{-1}\mathcal{L}$ with a
suitable operator $\mathcal{M}^{-1}$ which should be as close to
the pseudoinverse of $\mathcal{L}$ as possible in order to
separate the zero eigenvalue from the rest of the spectrum of
$\mathcal{L}$ and thus speed up the convergence.\cite{eirola} Of
course, in practice one does not have a routine for a
pseudoinverse of $\mathcal{L}$ and some heuristic preconditioning
must be used. We used as the preconditioning the inverse of the
``Sylvester part" $\mathcal{L}_0$ of $\mathcal{L}$. If we write
$\mathcal{L}\hat{\rho}=\hat{A}\hat{\rho}+\hat{\rho}\hat{A}^{\dag}
+\sum_i \hat{B}_i \hat{\rho}\,\hat{B}_i^{\dag}$ then the Sylvester
part is given by
$\mathcal{L}_0\hat{\rho}=\hat{A}\hat{\rho}+\hat{\rho}\hat{A}^{\dag}$.
Performing the inversion $\mathcal{L}_0^{-1}$ amounts to solving
the Sylvester equation which is a relatively fast procedure
scaling with $N^3$. The usage of the preconditioning was in our
case crucial for the convergence. After the iteration reaches its
end the stationary density matrix is obtained by imposing the
unity trace condition to the solution, i.e.\
$\mathbf{x}\leftrightarrow \hat{\rho}^{\rm stat},\
\mathrm{Tr}_{\rm sys}\hat{\rho}^{\rm stat}=1$.

The next step is to calculate the zero-frequency current noise
from \eqref{noise-general}, \eqref{gmres}. The equation
\eqref{gmres} can be solved iteratively in the Krylov subspace by
the {\em generalized minimum residual} method (GMRes). If
$\mathbf{x}_0$ is an initial approximation for the solution of
$\mathcal{L}\mathbf{x}=\mathbf{b}$ the Krylov subspace is
generated by the Arnoldi iteration starting with the vector
$\mathbf{r}_0=\mathbf{b}-\mathcal{L}\mathbf{x}_0$ and the GMRes
method finds a vector $\mathbf{x}\in \mathbf{x}_0
+\mathcal{K}_j(\mathcal{L},\mathbf{r}_0)$ that minimizes the norm
of the residual $\mathbf{r}=\mathbf{b}-\mathcal{L}\mathbf{x}$. The
vector $\mathbf{x}$ is assumed in the form
$\mathbf{x}=\mathbf{x}_0+\mathbf{Q}_j\cdot\xi$ and the solution
that minimizes the norm of the residual is obtained from
\begin{widetext}
\begin{equation}
\begin{split}
&\min\,\norm{\mathbf{b}-\mathcal{L}\mathbf{x}}
=\min\,\norm{\mathbf{b}-\mathcal{L}(\mathbf{x}_0+\mathbf{Q}_j\cdot\xi)}
=\min\,\norm{\mathbf{r}_0-\mathcal{L}\mathbf{Q}_j\cdot\xi}
=\min\,\norm{\mathbf{r}_0-\mathbf{Q}_{j+1}\cdot\mathbf{H}_j\cdot\xi}
\\&=\min\,\norm{\mathbf{Q}_{j+1}\cdot(\mathbf{e}_1 \beta
-\mathbf{H}_j\cdot\xi)}
=\min\,\norm{\mathbf{e}_1\beta-\mathbf{H}_j\cdot\xi} ,\text{ with
} \beta=\norm{\mathbf{r}_0} \text{ and }
\mathbf{e}_1=(1,0,\dots,0)^T\ .
\end{split}
\end{equation}
\end{widetext}
The last minimization problem is solved easily by the {\em
QR-decomposition} of the small rectangular matrix
$\mathbf{H}_j=\mathbf{U}\mathbf{R}$, where
$\mathbf{U}\in\mathbb{C}^{(j+1)\times j}$ has orthonormal columns
($\mathbf{U}^{\dag}\mathbf{U}=\mathbf{I}$) and
$\mathbf{R}\in\mathbb{C}^{j\times j}$ is upper triangular. If
$\mathbf{H}_j$ has full rank the solution to the minimization
problem is obtained by solving
$\mathbf{R}\cdot\mathbf{\xi}=\beta\,\mathbf{U}^{\dag}\cdot\mathbf{e}_1$.
If $\norm{\mathbf{b}-\mathcal{L}\mathbf{x}} > tol$ the
$\mathbf{x}_0,\mathbf{r}_0$ are replaced by
$\mathbf{x},\mathbf{r}$ and the sequence of steps is restarted.
Again, the iteration may not converge without preconditioning. We
used the same preconditioning as in the calculation of the null
vector, i.e.\ we solved the problem
$\mathcal{L}_0^{-1}\mathcal{L}\mathbf{x}=\mathcal{L}_0^{-1}\mathbf{b}$
by the above described algorithm. In the end of the iteration we
fixed the solution by removing any component in the direction of
the null vector by imposing the trace condition of \eqref{gmres}.

It has to be noted that the choice of some suitable
preconditioning is the difficult part of the problem and most
probably there is no general hint how to proceed. Particular cases
must be attempted anew based on experience and intuition. For
example, we tried to solve our model for some parameters with the
damping kernel \eqref{damping} replaced by its translationally
invariant form from Ref.~\onlinecite{nov-prl-03}. The same
preconditioning yielded a convergent iteration scheme in much
restricted range of the device biases compared to the rotating
wave approximation form of the damping used otherwise. Also the
non-zero temperature calculations converged significantly slower
than the corresponding zero-temperature counterparts. In the
sequential tunneling limit the non-zero temperature calculations
actually failed to converge at all so that the data presented in
Fig.~\ref{ST_results} had to be calculated with a direct method.
Fortunately, the oscillator is in that limit close to its
equilibrium state so that we needed $N=15$ at maximum which made
the direct calculations feasible.

As for the implementation of the numerical algorithms we used
MATLAB on personal computers and/or Linux workstations. The
building blocks are handy in MATLAB including the preconditioned
GMRes routine with restarts which solves completely the noise
calculation part of the problem. For efficiency reasons the
stationary part of the code was written ``from the scratch" within
MATLAB. The memory requirements were negligible (about 10-20 MB of
RAM for $N$ up to 40) and the calculation for $N=40,\, T=0$ for a
given set of the other parameters lasted a few minutes on a Linux
workstation, moderately depending on the parameters via the number
of required iterations to reach the convergence (a factor of
2--3). As already mentioned the non-zero temperature calculations
were much slower and could take up to an hour for a given set of
parameters. Most of the calculations were done for $N=25$, though,
since this level of truncation was usually sufficient as tested by
comparing results with different values of $N$. We also checked
occasionally that different choices of junctions for the
calculation of the mean current and the noise
\eqref{noise-general} gave the same numerical results within a
very high accuracy.

\section{Microscopic derivation of the rate equation}
\label{P(E)-theory}

In this appendix we give the derivation of the rate equation
describing the sequential tunneling regime realized in the limit
of the weak inter-dot coupling $V_0\to 0$. To this end we solve
the $n$-unresolved version of the GME \eqref{GME1} using the
lowest order perturbation theory in $V_0$. For small $V_0$ the
rates (proportional to $V_0^2$) are small and we may assume that
the oscillator gets equilibrated between individual tunneling
events between the adjacent dots. Within these assumptions we can
find a closed set of equations for only the occupations of the
respective dots $P_L,\,P_C,\,P_R$ plus the probability that the
device is empty $P_0$ ($P_L+P_C+P_R+P_0=1$).

These quantities defined as $P_I=\langle I|
\mathrm{Tr_{osc,B}}\hat{\sigma}|I\rangle\ (I=0,L,R,C)$ obey the
following equations stemming from \eqref{GME1}
\begin{equation}\label{continuity}
\begin{split}
\dot{P}_0 &= -\Gamma P_0 + \Gamma P_R\\
\dot{P}_L &= \Gamma P_0
             +i\mathrm{Tr_{osc,B}}(\hat{\sigma}_{LC}t_L(\hat{x})-t_L(\hat{x})\hat{\sigma}_{CL})\\
          &=\Gamma P_0-2\mathrm{Im}[\mathrm{Tr_{osc,B}}(\hat{\sigma}_{LC}t_L(\hat{x}))]\\
\dot{P}_C &=i\mathrm{Tr_{osc,B}}(\hat{\sigma}_{CL}t_L(\hat{x})-t_L(\hat{x})\hat{\sigma}_{LC}\\
          &\qquad+\hat{\sigma}_{CR}t_R(\hat{x})-t_R(\hat{x})\hat{\sigma}_{RC})\\
          &=2\mathrm{Im}[\mathrm{Tr_{osc,B}}(\hat{\sigma}_{LC}t_L(\hat{x}))]
          +2\mathrm{Im}[\mathrm{Tr_{osc,B}}(\hat{\sigma}_{RC}t_R(\hat{x}))]\\
\dot{P}_R &= -\Gamma P_R+i\mathrm{Tr_{osc,B}}(\hat{\sigma}_{RC}t_R(\hat{x}))-t_R(\hat{x})\hat{\sigma}_{CR}\\
          &=-\Gamma P_R-2\mathrm{Im}[\mathrm{Tr_{osc,B}}(\hat{\sigma}_{RC}t_R(\hat{x}))]
          \ .
\end{split}
\end{equation}
We notice explicitly that the charge (probability) conservation
condition $\dot{P}_0+\dot{P}_L+\dot{P}_C+\dot{P}_R=0$ is
fulfilled. The occupations couple to the off-diagonal elements
$\hat{\sigma}_{LC},\,\hat{\sigma}_{CR}$ satisfying
\begin{align}
\dot{\hat{\sigma}}_{LC} &= -i\Bigl(\frac{\eps_b}{2}
   \hat{\sigma}_{LC}+\hat{\sigma}_{LC}\frac{\eps_b}{2x_0}\,\hat{x}
   + [\hat{H}'_{\mathrm{osc}},\hat{\sigma}_{LC}]\Bigr)\notag\\
   &+i\bigl(\hat{\sigma}_{LL}t_L(\hat{x})-t_L(\hat{x})\hat{\sigma}_{CC}\bigr)
   +i\hat{\sigma}_{LR}t_L(\hat{x})\label{sigmaLC}\\
\dot{\hat{\sigma}}_{CR} &= -i\Bigl(-\frac{\eps_b}
    {2x_0}\,\hat{x}\,\hat{\sigma}_{CR} + \hat{\sigma}_{CR}\frac{\Delta
    V}{2}+[\hat{H}'_{\mathrm{osc}},\hat{\sigma}_{CR}]\Bigr)\notag\\
    &+i\bigl(\hat{\sigma}_{CC}t_R(\hat{x})-t_R(\hat{x})\hat{\sigma}_{RR}\bigr)
    -i t_L(\hat{x})\hat{\sigma}_{LR}-\frac{\Gamma}{2}\,\hat{\sigma}_{CR} \ .
\end{align}
In the full generality, these equations would generate an infinite
hierarchy of equations for different moments of the whole density
matrix $\hat{\sigma}$. However, in the lowest order in $V_0$ we
can neglect the coupling to $\hat{\sigma}_{LR}$ (which is of
higher order in $V_0$) and formally integrate the equations
leading to
\begin{equation}
\begin{split}
\hat{\sigma}_{LC}(t)&=-i\int_0^{\infty}\!\!\!\!d\tau
\Big[e^{-i(\hat{H}'_{\mathrm{osc}}+\frac{\eps_b}
{2})\tau}t_L(\hat{x})\hat{\sigma}_{CC}(t-\tau)
e^{i(\hat{H}'_{\mathrm{osc}}-\frac{\eps_b}
{2x_0}\,\hat{x})\tau}\Big] \\
&+i\int_0^{\infty}\!\!\!\!d\tau
\Big[e^{-i(\hat{H}'_{\mathrm{osc}}+\frac{\eps_b}
{2})\tau}\hat{\sigma}_{LL}(t-\tau)t_L(\hat{x})e^{i(\hat{H}'_{\mathrm{osc}}
-\frac{\eps_b }{2x_0}\,\hat{x})\tau}\Big]
\end{split}
\end{equation}
and similarly for $\hat{\sigma}_{CR}(t)$. Now, we can employ the
standard Born-Markov approximation assuming the oscillator plus
bath subsystem in local equilibrium corresponding to a given
charge state, and neglecting the memory effects in the evolution
of $P_I(t)$'s (both assumptions are justified by the small $V_0$):
\begin{equation}\label{decoupling}
\begin{split}
\hat{\sigma}_{LL}(t-\tau)&\simeq\hat{\sigma}_{\mathrm{osc,B}}(0) P_L(t)\\
\hat{\sigma}_{CC}(t-\tau)&\simeq\hat{\sigma}_{\mathrm{osc,B}}
\big(\tfrac{\eps_b}{2x_0}\big) P_C(t)
\end{split}
\end{equation}
with $\hat{\sigma}_{\mathrm{osc,B}}(\lambda)=
e^{-\beta(\hat{H}'_{\mathrm{osc}}-\lambda\hat{x})}/Z(\lambda),\
Z(\lambda)=
\mathrm{Tr}_{\mathrm{osc,B}}(e^{-\beta(\hat{H}'_{\mathrm{osc}}-\lambda\hat{x})})$,
where $\mathrm{Tr}_{\mathrm{osc,B}}$ means tracing over the
oscillator and the heat bath.

The rate equations for the evolution of the probabilities are
thus:
\begin{equation}\label{rate_eq}
\begin{split}
\dot{P}_0 &= -\Gamma
P_0 + \Gamma P_R\\
\dot{P}_L &= \Gamma P_0- \Gamma_{CL}P_L+\Gamma_{LC}P_C \\
\dot{P}_C &= \Gamma_{CL}P_L-(\Gamma_{LC}+\Gamma_{RC})P_C+
\Gamma_{CR}P_R\\
\dot{P}_R &=\Gamma_{RC}P_C-(\Gamma_{CR}+\Gamma)P_R\\
\end{split}
\end{equation}
where the $\Gamma_{IJ}$'s, the transition rates from the state $J$
to $I$, are given by
\begin{widetext}
\begin{equation}\label{rates}
\begin{split}
\Gamma_{CL}&=2\mathrm{Re}\Big[\int_0^{\infty}\!\!\!\!d\tau
e^{-i\frac{\eps_b}{2}\tau}
\mathrm{Tr}_{\mathrm{osc,B}}\Big(e^{-i\hat{H}'_{\mathrm{osc}}\tau}
\hat{\sigma}_{\mathrm{osc,B}}(0)t_L(\hat{x})e^{i(\hat{H}'_{\mathrm{osc}}
-\frac{\eps_b }{2x_0}\,\hat{x})\tau} t_L(\hat{x})\Big)\Big]\\
\Gamma_{LC}&= 2\mathrm{Re}\Big[\int_0^{\infty}\!\!\!\!d\tau
e^{-i\frac{\eps_b}{2}\tau}
\mathrm{Tr}_{\mathrm{osc,B}}\Big(e^{-i\hat{H}'_{\mathrm{osc}}\tau}
t_L(\hat{x})\hat{\sigma}_{\mathrm{osc,B}}\big(\tfrac{\eps_b}{2x_0}\big)
e^{i(\hat{H}'_{\mathrm{osc}}-\frac{\eps_b}{2x_0}\,\hat{x})\tau} t_L(\hat{x})\Big)\Big]\\
\Gamma_{RC} &= 2\mathrm{Re}\Big[\int_0^{\infty}\!\!\!\!d\tau
e^{-\frac{\Gamma}{2}\tau}e^{i\frac{\eps_b}{2}\tau}
\mathrm{Tr}_{\mathrm{osc,B}}\Big(e^{-i\hat{H}'_{\mathrm{osc}}\tau}t_R(\hat{x})
\hat{\sigma}_{\mathrm{osc,B}}\big(\tfrac{\eps_b}{2x_0}\big)
e^{i(\hat{H}'_{\mathrm{osc}}-\frac{\eps_b }{2x_0}\,\hat{x})\tau} t_R(\hat{x})\Big)\Big]\\
\Gamma_{CR} &= 2\mathrm{Re}\Big[\int_0^{\infty}\!\!\!\!d\tau
e^{-\frac{\Gamma}{2}\tau}e^{i\frac{\eps_b}{2}\tau}
\mathrm{Tr}_{\mathrm{osc,B}}\Big(e^{-i\hat{H}'_{\mathrm{osc}}\tau}
\hat{\sigma}_{\mathrm{osc,B}}(0)t_R(\hat{x})e^{i(\hat{H}'_{\mathrm{osc}}
-\frac{\eps_b }{2x_0}\,\hat{x})\tau} t_R(\hat{x})\Big)\Big]\ .
\end{split}
\end{equation}
\end{widetext}
These rates can be also obtained starting from the Fermi Golden
Rule expression for the bath-assisted electronic transitions
($P(E)$-theory\cite{nazarov}) bearing in mind that the electronic
state on the right dot is broadened by $\tfrac{\Gamma}{2}$ due to
the coupling to the (empty) right lead which causes the appearance
of the $e^{-\frac{\Gamma}{2}\tau}$ factors in the expressions for
$\Gamma_{RC},\Gamma_{CR}$.

To evaluate the rates we generalize the method used by Braig and
Flensberg\cite{braig-prb-03} for the $\alpha=0$ case. The shifted
Hamiltonian $\hat{H}'_{\mathrm{osc}}-\frac{\eps_b}{2x_0}\hat{x}$
can be eliminated by performing a suitable unitary transformation
which is a generalization of the well-known polaron shift from the
independent boson model \cite{mahan} to more oscillator modes and
which is given by the unitary operator \cite{braig-prb-03}
\begin{equation}\label{unitary}
\hat{S} = e^{-i\hat{A}}, \hat{A}=\hat{p}\,l+\sum_j \hat{p}_j\,l_j
\end{equation}
where $l$ and $l_j$ are constants to be determined so that the
linear shift is cancelled. It was found in
Ref.~\onlinecite{braig-prb-03} that
\begin{align}
l=\frac{-\eps_b}{2x_0 m\omega_0^2},\
l_j=\frac{c_j\,l}{m_j\omega_j^2}\label{shift} \\
\intertext{and}
  \hat{H}'_{\mathrm{osc}}-\frac{\eps_b}{2x_0}\hat{x}
  =\hat{S}^{\dagger}\hat{H}'_{\mathrm{osc}}\hat{S}-\frac{\eps_b^2}{8x_0^2m\omega_0^2}\ .
\end{align}

We may thus rewrite the expression for, e.g., the $\Gamma_{CL}$
rate as
\begin{equation}
\begin{split}
\Gamma_{CL}&=2\mathrm{Re}\Big[\int_0^{\infty}d\tau
e^{-i\frac{\eps_b}{2}(1+\frac{\eps_b}{4x_0^2m\omega_0^2})\tau}\\
&\big\langle
e^{-i\hat{H}'_{\mathrm{osc}}\tau}t_L(\hat{x})\hat{S}^{\dagger}
e^{i\hat{H}'_{\mathrm{osc}}\tau}\hat{S}t_L(\hat{x})\big\rangle_0\Big],
\end{split}
\end{equation}
with the expectation value $\langle\,\bullet\,\rangle_0
=\mathrm{Tr}_{\mathrm{osc,B}}(\,\bullet\
\hat{\sigma}_{\mathrm{osc,B}}(0))$. Using the Baker-Hausdorff
theorem and introducing the function $F(\tau ;\alpha) =\big\langle
e^{i\hat{A}(\tau)-\alpha\hat{x}(\tau)}e^{-i\hat{A}-\alpha\hat{x}}\big\rangle_0$
satisfying $F^{\ast}(\tau;\alpha)=F(-\tau;\alpha)$ we get
\begin{subequations}\label{rates_F_function}
\begin{equation}
\Gamma_{CL}=V_0^2e^{ -2\alpha(x_0-l/2)}\tilde{F}\big(\omega =
\tfrac{\eps_b}{2}(1+\tfrac{\eps_b}{4x_0^2m\omega_0^2});\,\alpha\big)\
.
\end{equation}
Similarly, for the corresponding backward rate $\Gamma_{LC}$ we
get
\begin{equation}
\Gamma_{LC}=V_0^2 e^{-2\alpha(x_0-l/2)}\tilde{G}\big(\omega =
-\tfrac{\eps_b}{2}
(1+\tfrac{\eps_b}{4x_0^2m\omega_0^2});\,\alpha\big)\ ,
\end{equation}
with the function $G(\tau ;\alpha) = \big\langle
e^{-i\hat{A}(\tau)-\alpha\hat{x}(\tau)}
e^{i\hat{A}-\alpha\hat{x}}\big\rangle_0$. The transfer rates
between the central and right dot read
\begin{align}
\Gamma_{RC}&=V_0^2e^{-2\alpha(x_0+l/2)}\notag\\
\times&\int_{-\infty}^{\infty}\frac{d\omega}{2\pi}\,
\tilde{F}(\omega;\alpha)\frac{\Gamma}{(\omega-\tfrac{\eps_b}{2}
(1-\frac{\eps_b}{4x_0^2m\omega_0^2}))^2+(\frac{\Gamma}{2})^2}\ , \label{cr_rate}\\
\Gamma_{CR}&=V_0^2e^{-2\alpha(x_0+l/2)}\notag\\
\times&\int_{-\infty}^{\infty}\frac{d\omega}{2\pi}\,
\tilde{G}(\omega;\alpha)\frac{\Gamma}{(\omega+\frac{\eps_b}{2}(1-\frac{\eps_b}{4x_0^2m\omega_0^2}))^2+(\frac{\Gamma}{2})^2}\
.
\end{align}
\end{subequations}

The evaluation of the functions $\tilde{F}(\om;\alpha)$ and
$\tilde{G}(\om;\alpha)$ follows a standard route found in
textbooks (Ref.~\onlinecite{mahan}, Sec.\ 4.3;
Ref.~\onlinecite{gardiner}, Sec.\ 4.4), or
Ref.~\onlinecite{kittel}, Ch.\ 20). Technically, the task is to
evaluate a particular characteristic function of a
(multidimensional) Gaussian distribution. The result is again
Gaussian, into which only second-order correlation functions
enter.

We introduce the operator $\hat{\tilde{A}}(\tau;
\alpha)=\hat{A}(\tau)+i\alpha\hat{x}(\tau)$, so that $F(\tau;
\alpha)= \big\langle e^{i\hat{\tilde{A}}(\tau; \alpha) }
e^{-i\hat{\tilde{A}}^{\dagger}(0; \alpha)}\big\rangle_0$ and
$G(\tau; \alpha)= \big\langle e^{-i\hat{\tilde{A}}(\tau; -\alpha)
}e^{i\hat{\tilde{A}}^{\dagger}(0; -\alpha)}\big\rangle_0$. Since
$\hat{H}'_{\mathrm{osc}}$ is quadratic in $\hat{x},\, \hat{x}_j$
and $\hat{p},\, \hat{p}_j$ we may rewrite $F$ and $G$ as $F(\tau
;\alpha) = \exp\big(\tfrac{1}{2}\big\langle 2\hat{\tilde{A}}(\tau
;\alpha)\hat{\tilde{A}}^{\dagger}(0;\alpha)-\hat{\tilde{A}}^2(\tau;\alpha)
-\hat{\tilde{A}}^{\dagger 2}(0;\alpha)\big\rangle_0\big),\,G(\tau
;\alpha) = \exp\big(\frac{1}{2}\big\langle 2\hat{\tilde{A}}(\tau
;-\alpha)\hat{\tilde{A}}^{\dagger}(0;-\alpha)
-\hat{\tilde{A}}^2(\tau;-\alpha)-\hat{\tilde{A}}^{\dagger
2}(0;-\alpha)\big\rangle_0\big)$ and we have thus established that
\begin{equation}\label{FG_functions}
 G(\tau;\alpha)=F(\tau;-\alpha)\ .
\end{equation}
The function $F(\tau; \alpha)$ can be rewritten in terms of the
following auxiliary quantity ($\hat{A}\propto l$, see Eqs.\
\eqref{unitary}, \eqref{shift})
\begin{equation}
\begin{split}
E(\tau ;\alpha;l) &= \big\langle\hat{\tilde{A}}(\tau;\alpha )
\hat{\tilde{A}}^{\dagger}(0;\alpha)\big\rangle_0 \\
&=\big\langle\big(\hat{A}(\tau)+i\alpha\hat{x}(\tau)\big)\big(\hat{A}(0)-i\alpha
x(0)\big)\big\rangle_0 \ .
\end{split}
\end{equation}

We evaluate $E(\tau;\alpha;l)$ following the lines of
Ref.~\onlinecite{braig-prb-03} where $E(\tau;0;l)$ was evaluated.
The idea is to express the function $E$ in terms of the retarded
Green's function
\begin{equation}
E^R(\tau;\alpha;l)=-i\theta(\tau)\big\langle[\hat{\tilde{A}}(\tau),
\hat{\tilde{A}}^{\dagger}(0)]\big\rangle_0,
\end{equation}
using the fluctuation-dissipation theorem
\begin{equation}
\tilde{E}(\omega;\alpha;l)=-2\mathrm{Im}\big(\tilde{E}^R(\omega;\alpha;l)\big)\big(1+n_B(\omega)\big)
\end{equation}
and then find $E^R$ by solving its equation of motion in the
Fourier space (for details of the derivation see
Ref.~\onlinecite{christian}). Assuming the Ohmic coupling between
the oscillator and the heat bath, i.e.\ $J(\om)= m\gamma\om
f(\tfrac{\om}{\om_c})$, we find
\begin{widetext}
\begin{equation}
\begin{split}
\tilde{E}^R(\omega; \alpha;l)&=\frac{m\omega_0^2}
{\omega^2-\omega^2_0+i\gamma\omega}\Big[l^2\Big(1+i\frac{\gamma}{\omega}\Big)
-\frac{2\alpha l \om}
{m\omega_0^2}\Big(1+i\frac{\gamma}{\omega}\Big)+\frac{\alpha^2}{m^2\om_0^2}\Big]\
,
\end{split}
\end{equation}
which coincides with the result of Ref.~\onlinecite{braig-prb-03}
for $\alpha=0$. We finally arrive at the expression for the $F$
function
\begin{equation}\label{F_function}
\begin{split}
F(\tau ; \alpha )
&=\exp\Big[\int_{-\infty}^{\infty}\frac{d\omega}{2\pi}\Big(\tilde{E}(\omega;\alpha;l)
\,e^{-i\omega\tau}-\tilde{E}(\omega;0;l)+ \tilde{E}(\omega ;\alpha;0)\Big)\Big]\\
&=\exp\Big[\int_{-\infty}^{\infty}\frac{d\omega}{\pi}\frac{1+n_B(\omega)}{\omega}
\frac{m\omega_0^2\gamma}{(\omega^2-\omega_0^2)^2+\gamma^2\omega^2}
\Big(\big(l^2\omega_0^2-\frac{2\alpha
l\omega}{m}+\frac{\alpha^2\omega^2}{m^2\omega_0^2}\big)\,e^{-i\omega\tau}-
l^2\omega_0^2+\frac{\alpha^2\omega^2}{m^2\omega_0^2}\Big)\Big]\ .
\end{split}
\end{equation}
\end{widetext}
The analytical structure of the $F$ function, in particular the
power law decay for large times at zero temperature
$F(\tau)\propto \tau^{-\delta},\, \tau\to\infty,\,
T=0,\,\delta=\tfrac{ml^2\gamma}{\hbar\pi}$, remains the same as in
the $\alpha=0$ case \cite{braig-prb-03} since it only depends on
the behavior of the prefactor $l^2\omega_0^2-\tfrac{2\alpha
l\omega}{m}+\tfrac{\alpha^2\omega^2}{m^2\omega_0^2}$ at $\om\to
0^+$.

\bibliography{noise}

\end{document}